\documentclass[aps,prb,twocolumn,groupaddress]{revtex4}

\usepackage[dvips]{graphicx}
\usepackage{color}
\usepackage{CJK}

\renewcommand{\vec}[1]{\mathbf{#1}}

\begin{document}
\begin{CJK*}{GB}{gbsn}

\title{Experimental Phase Diagram and Dynamics of a Dilute Dipolar-Coupled Ising System}

\author{J.~A.~Quilliam}\altaffiliation{Present address: Laboratoire de Physique des Solides, Universit\'{e} Paris-Sud 11, UMR CNRS 8502, 91405 Orsay, France } 
\author{S.~Meng (ÃÏÊ÷³¬)} 
\author{J.~B.~Kycia}
\affiliation{Department of Physics and Astronomy, Guelph-Waterloo Physics Institute and Institute for Quantum Computing, University of Waterloo, Waterloo, ON N2L 3G1 Canada}

\date{\today}

\begin{abstract}
We present ac susceptibility and specific heat measurements taken on samples of LiHo$_x$Y$_{1-x}$F$_4$ in the dilute limit: $x = 0.018$, 0.045, 0.080 and 0.12.  Susceptibility measurements show glassy behavior including wide absorption spectra that continually broaden with decreasing temperature.  Dynamical scaling analyses show evidence of finite-temperature spin-glass transitions, the temperatures of which match those of recent theoretical work.  A surprisingly long intrinsic time constant is observed in these samples and is found to be inversely correlated with the concentration of magnetic moments, $x$.  Our results support the picture that this behavior is largely a single-ion effect, related to the random transverse fields generated by the off-diagonal component of the dipolar interaction and significantly slowed by the important nuclear hyperfine interaction.  Specific heat measurements show broad features due to the electronic spins on top of a large Schottky-like nuclear contribution.  Unusually, the peak position of the electronic component is found to be largely concentration independent, unlike the glass transition temperature.
\end{abstract}


\maketitle
\end{CJK*}

\section{Introduction}

The compound LiHoF$_4$ and derivatives obtained through dilution of the magnetic Ho$^{3+}$ moments with non-magnetic Y$^{3+}$, have long been considered to be excellent representations of the dipolar Ising model and have been studied extensively as model magnets.\cite{Reich1990,Bitko1996}  In recent years, however, this series of materials has become the subject of considerable debate, both experimental\cite{Quilliam2007,Jonsson2007,Quilliam2008,AnconnaTorres2008,Jonsson2008} and theoretical,\cite{Biltmo2007,Biltmo2008,Tam2009} particularly regarding the dilute (small $x$) limit and the question of whether the introduction of randomness and frustration originating from the dipolar interaction leads to a spin-glass transition or something more exotic.\cite{Ghosh2002,Ghosh2003}

There are two main components to this debate.  First is the largely theoretical question of whether the simple toy model that is thought to describe LiHo$_x$Y$_{1-x}$F$_4$, the dilute, dipolar-coupled Ising model, ought to exhibit a finite-temperature spin-glass transition or not.  While mean field theory suggests that there is a spin-glass transition,\cite{Stephen1981} some Monte Carlo simulations have called that conclusion into question.\cite{Snider2005,Biltmo2007,Biltmo2008}  However, recent Monte Carlo results, implementing parallel tempering have shown strong evidence of a finite-temperature spin-glass transition.\cite{Tam2009}

The other major part of the debate deals with the real material and the possibility that the toy model is in fact not an adequate description of the system's underlying model.  Several experimental groups have come to different conclusions regarding the dilute regime.  One research group has come to the conclusion that at $x = 0.167$ and $x=0.20$, the materials undergo a spin-glass transition at 130 mK and 150 mK, respectively, as determined from the temperature dependence of the ac susceptibility $\chi(\omega)$ and nonlinear susceptibility $\chi_3$,\cite{Reich1990,Wu1991,Wu1993,AnconnaTorres2008} whereas at $x = 0.045$ an unusual spin liquid state appears.\cite{Reich1987,Reich1990,Ghosh2002,Ghosh2003,Silevitch2007prl}  This spin liquid or ``antiglass'' state is primarily characterized by a narrowing of the absorption spectrum $\chi''(\omega)$ as the temperature is lowered.\cite{Reich1987,Reich1990,Ghosh2002}  Several other interesting effects were observed in the 4.5\% material including sharp features in the specific heat at 120 and 300 mK,~\cite{Reich1990,Ghosh2002} contrasting with a smooth $T^{-0.75}$ power law dependence of the dc susceptibility,\cite{Ghosh2002} hole burning in the absorption spectrum and possible coherent oscillations.\cite{Ghosh2003}

A second experimental group, in contrast, has found qualitatively similar behavior between 4.5\% and 16.7\% samples.\cite{Jonsson2007,Jonsson2008}  J\"{o}nsson \emph{et al.}~\cite{Jonsson2007,Jonsson2008} have concluded that neither sample shows a sufficiently diverging $\chi_3$ to be considered a spin glass with a finite-temperature glass transition, $T_g$, instead suggesting that the materials are a kind of superparamagnet with thermally activated dynamics.

We present here, and in other short articles,~\cite{Quilliam2007,Quilliam2008} a third point of view: that there is a finite-temperature spin-glass transition in samples of LiHo$_x$Y$_{1-x}$F$_4$ for $x\lesssim 0.2$, and that there is \emph{no} exotic antiglass state at $x = 0.045$.  We have performed ac susceptibility measurements using a specially designed SQUID magnetometer on samples of $x = 0.018$, 0.045 and 0.080 and have performed specific heat measurements on those same samples plus an additional $x = 0.12$ sample.  The data present several puzzles and conclusions are hampered in this system by extremely long time constants, even well above $T_g$, which can be seen to be a result of single ion physics and the importance of the nuclear hyperfine interaction.~\cite{Atsarkin1988,Schechter2008b}  Nonetheless, the evidence largely supports the picture of spin glass physics in these materials.

This article begins with a discussion of the microscopic Hamiltonian that is thought to describe the LiHo$_x$Y$_{1-x}$F$_4$ system and a review of past work on the phase diagram of this series of materials.  We concentrate on the low-$x$ part of the phase diagram, but refer the reader to the recent and more inclusive review article Ref.~\onlinecite{Gingras2011}.   This is followed by a description of the apparatus and methods used in this work, specifically, our specific heat and ac susceptibility experiments.  In this work, we expand on previous measurements of the specific heat~\cite{Quilliam2007} and ac susceptibility~\cite{Quilliam2008} of LiHo$_x$Y$_{1-x}$F$_4$, covering a larger range of $x$.  Specifically, the new work contained within this article consists of ac susceptibility measurements on $x=0.018$ and $x=0.080$ samples as well as a specific heat measurement at $x=0.12$.   In Section IV, we present our results along with scaling analyses that are strongly suggestive of spin glass physics. An explanation~\cite{Atsarkin1988,Schechter2008b} for some of the dynamical behavior that is observed, based on the single-ion Hamiltonian of the system, is presented in Section V.  Our results are then compared with those of other groups and disparities in data and interpretation are discussed.

\section{Background}

		
\begin{figure}
\begin{center}
\includegraphics[width=2.7in,keepaspectratio=true]{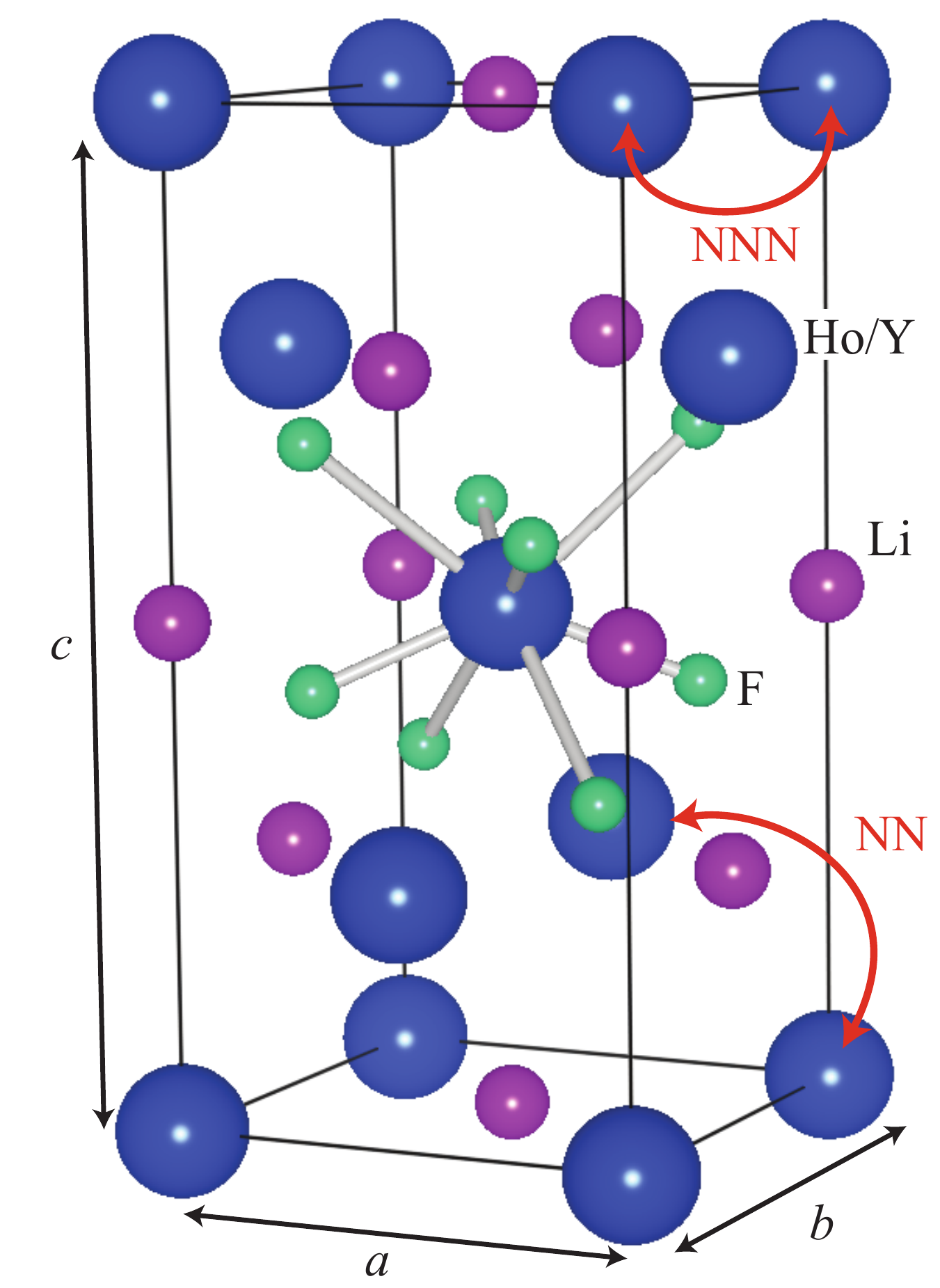}
\caption{(color online) The crystal structure of LiHoF$_4$.  The Ho$^{3+}$ moments may also be randomly replaced by non-magnetic Y$^{3+}$ ions during crystal growth.  Only the F$^-$ ions immediately surrounding one of Ho$^{3+}$/Y$^{3+}$ sites have been drawn in order to simplify the image.  The crystal structure provides a  crystal field with $S_4$ symmetry around the Ho$^{3+}$ ions, leading to an Ising ground-state doublet.   On dilution, random frustration is provided by the anisotropy of the dipolar interaction.  The dipolar interaction is ferromagnetic between nearest neighbors (NN) but antiferromagnetic between next nearest neighbors (NNN).
\label{CrystalStructure}
}
\end{center}
\end{figure}

The magnetic Ho$^{3+}$ ions in LiHo$_x$Y$_{1-x}$F$_4$ possess spin $S=2$, orbital angular momentum $L=6$ and total angular momentum $J=8$ moments as a result of Hund's rules and strong spin-orbit coupling, typical of rare earth ions. The $J=8$ levels are further split in energy by the crystalline electric field, of $S_4$ symmetry.  The resulting Hamiltonian can be expressed in terms of Steven's operators,\cite{Hutchings1964} as
\begin{equation}
\mathcal{H}_\mathrm{CF} = \sum_{m,\alpha} B_m^\alpha O_m^\alpha.
\end{equation}
Ho$^{3+}$ is a non-Kramers ion, so it is only by virtue of the symmetry of the crystal field, that an Ising doublet ground state results.  This ground-state doublet, consisting of states $|\uparrow \rangle$ and $|\downarrow\rangle$, has no matrix elements of $J_x$ or $J_y$.  It does, however, exhibit large moments along the $z$-direction ($c$-axis).
\begin{equation}
\langle \uparrow | J_z | \uparrow \rangle = - \langle \downarrow | J_z | \downarrow \rangle = 5.15
\end{equation}
thus the effective $g$-factor is 
\begin{equation}
g_\mathrm{eff} = 2g_J \langle J_z \rangle = 13.8
\end{equation}
where $g_J = 5/4$ is the Ho$^{3+}$ Land\'{e} $g$-factor.  The next excited state of the Ho$^{3+}$ ions, $|\gamma\rangle$, has been measured to be roughly 11 K in energy above the Ising doublet by several groups in the past.\cite{Ronnow2007,Hansen1975}  Thus for the temperatures of interest, say near and below $T_c(x=1) = 1.53$ K,\cite{Hansen1975,Cooke1975,Beauvillain1978} this system is very well described as an Ising magnet.  

Because the $4f$-electrons of the rare-earth ions are tightly bound, they tend to possess relatively small exchange interactions.  When this is combined with the large magnetic moments, the dipole-dipole interaction can become energetically quite important.  In LiHo$_x$Y$_{1-x}$F$_4$, the dipolar interaction is in fact dominant; the nearest neighbor exchange interaction has been estimated experimentally to be roughly half the nearest neighbor dipolar interaction\cite{Mennenga1984} and in a recent theoretical work was determined to be a factor of 10 smaller.\cite{Biltmo2007}  The dipole-dipole Hamiltonian is given as 
\begin{equation}
\label{HamD}
\mathcal{H}_D = \sum_{\langle i,j \rangle} \frac{\mu_0}{4\pi} g_J^2 \mu_B^2
\left[ \frac{\vec{J}_i\cdot \vec{J}_j}{r_{ij}^3} - \frac{3(\vec{J}_i\cdot \vec{r}_{ij})(\vec{J}_j\cdot \vec{r}_{ij})}{r_{ij}^5} \right].
\end{equation}
Within the Ising doublet manifold, however, it can be written instead as
\begin{equation}
\mathcal{H}_D = \sum_{\langle i,j \rangle} \frac{\mu_0}{4\pi} g_\mathrm{eff}^2 \mu_B^2 \left( \frac{r_{ij}^2 - 3z_{ij}^2}{r_{ij}^5} \right) S_i^z S_j^z
\end{equation}
where the $S_i^z$ are spin-1/2 operators and 
\begin{equation}
g_\mathrm{eff} = 2g_J\langle \uparrow | J^z | \uparrow\rangle.
\end{equation}

This interaction has two important distinctions.  First, it is long range in nature, falling off as $1/r^3$.  This means that every spin is coupled to every other spin, at least to some degree, making a percolation threshold impossible, for example.  Second, the sign of the interaction is dependent on the angle of the vector connecting the spins and this always leads to a degree of frustration in three dimensions.  If the vector connecting spins is aligned with the Ising $c$-axis (so that they are stacked on top of each other in Fig.~\ref{CrystalStructure}), they are ferromagnetically coupled; if they are positioned within the same $ab$-plane (perpendicular to the Ising axis), they are antiferromagnetically coupled.  In LiHo$_x$Y$_{1-x}$F$_4$, nearest neighbors (NN) exhibit ferromagnetic coupling where next nearest neighbors (NNN) exhibit antiferromagnetic coupling, as shown in Fig.~\ref{CrystalStructure}, causing frustration.  The ferromagnetic couplings dominate in the pure material, which has a ferromagnetic transition temperature of 1.53 K.\cite{Hansen1975,Cooke1975,Beauvillain1978,Bitko1996} Upon randomly diluting the system with non-magnetic Y$^{3+}$, however, this frustration is exposed and is one of the main ingredients that gives rise to the glassy physics and possible spin glass state seen at $x \lesssim 0.25$.\cite{Reich1990}

Introduction of a magnetic field results in the Zeeman energy $\mathcal{H}_Z = g_J \mu_B \vec{H}\cdot \vec{J}$.  Much of the past research performed on this system has aimed at understanding the effect of quantum fluctuations, introduced with a magnetic field transverse to the $c$-axis, or $H_\perp$.  At the single-ion level, such a transverse field splits the Ising doublet by an energy of $\Delta(H_\perp) = 2\Gamma$ and mixes it with the next excited state $|\gamma\rangle$.  When projected onto a $S=1/2$ model,\cite{Tabei2008a,Chin2007} one has an effective transverse magnetic field of $\Gamma$ so the contribution $\Gamma S^x$.  The result is a manifestation of the famous transverse field Ising model (TFIM),\cite{Sachdev1999} which is one of the simplest models known to exhibit a zero-temperature, quantum phase transition (QPT).  
This paradigm was applied by Bitko \emph{et al.} to the ferromagnetic parent compound ($x=1$).\cite{Bitko1996} As $H_\perp$ is increased, quantum fluctuations become more and more powerful and eventually melt the ferromagnetic order, leading to a quantum paramagnetic state.  In LiHoF$_4$, this is seen to occur at a critical field $H_\perp^C = 4.9$ T.\cite{Bitko1996}  A number of theoretical works have attempted to fit the $(H_\perp,T)$ phase boundary, but a persistent quantitative disagreement has been found.\cite{Chakraborty2004,Tabei2008b}  The previous experimental phase boundary has recently been confirmed with dilatometry measurements,\cite{Dunn2010} and the mismatch between theory and experiment remains an outstanding problem.

The last component to the system's Hamiltonian to mention is the hyperfine coupling to the $I=7/2$ nuclear moment.  Holmium, in fact, has an extraordinarily large hyperfine interaction described by 
\begin{equation}
\mathcal{H}_\mathrm{HF} = A \vec{I}\cdot \vec{J}
\end{equation}
where $A = 40.21$ mK.\cite{Magarino1976,Magarino1980,Mennenga1984}  In the perfect Ising model, where no excited crystal field states are considered, this leads to 8 distinct electronuclear energy levels, separated by 207 mK.  While smaller than 1.53 K, thus not terribly influential in zero field for the parent compound, it is easy to see that the hyperfine coupling can become energetically important as one moves to low concentrations of $x < 0.25$, where glassy physics occurs.  The hyperfine interaction also has important consequences near the QPT of higher concentration samples, creating a non-trivial $(H_\perp,T)$ phase diagram\cite{Bitko1996,Chakraborty2004} and influencing the spectrum of excitations.\cite{Ronnow2005,Ronnow2007}

While a number of very interesting experiments have been performed on the ferromagnetic stoichiometries $x = 0.46$, $x = 0.67$ and $x = 1$,\cite{Bitko1996,Brooke1999,Brooke2001,Silevitch2007nature,Silevitch2010} we concentrate here on the dilute glassy portion of the phase diagram at $x = 0.2$ and below.  The first experiments performed on an $x = 0.167$ sample~\cite{Reich1990} showed largely conventional spin glass behavior.  Glassy relaxation was observed in the ac susceptibility, $\chi(f)$, and a broad feature was observed in the specific heat $C$.  Most evidence seemed to point to the existence of a dipolar spin-glass transition as was anticipated theoretically by Stephen and Aharony.\cite{Stephen1981}  LiHo$_x$Y$_{1-x}$F$_4$ is not the first system to be investigated as a possible dipolar spin glass.  Dilute Eu$x$Sr$_{1-x}$S was also considered as a candidate for such behavior, but is complicated by strong exchange interactions.\cite{Eiselt1979}  The low $x$ range of the LiHo$_x$Y$_{1-x}$F$_4$ series represents an important test of theories that either propose or deny the existence of a finite glass transition in a purely dipolar system.

\begin{figure*}
\begin{center}
\includegraphics[width=6in,keepaspectratio=true]{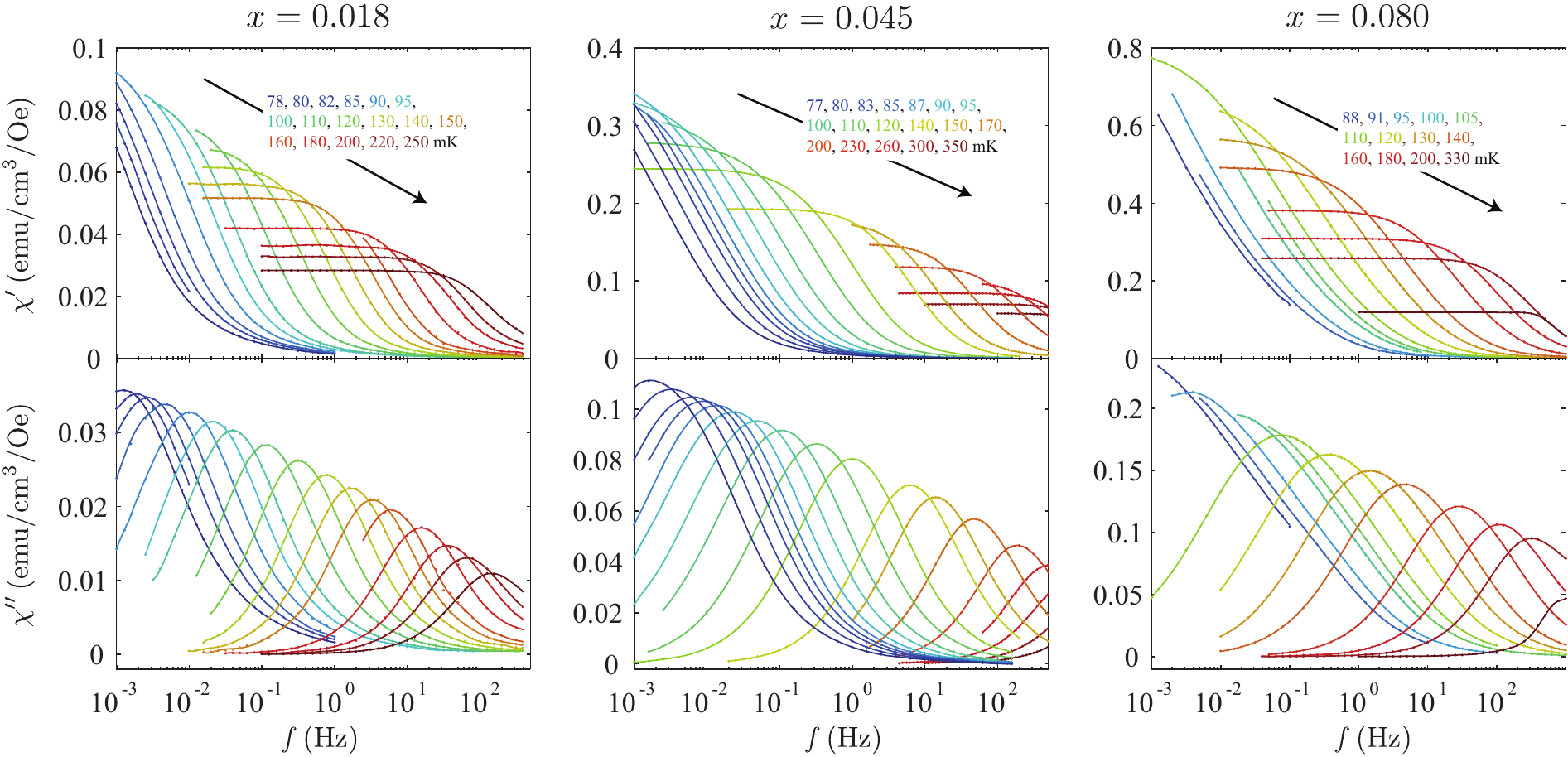}
\caption{(color online) Frequency scans of the ac susceptibility of LiHo$_x$Y$_{1-x}$F$_4$ for $x = 0.018$, $x=0.045$ and $x=0.08$.  The top panels show the in-phase susceptibility or $\chi'(f)$ while the bottom panels show the absorption spectra $\chi''(f)$. Temperature increases from left (blue) to right (red).  These data have been corrected for the demagnetization effect.
\label{FrequencyScans}
}
\end{center}
\end{figure*}

The effects of an applied transverse magnetic field were also tested at the $x=0.167$ stoichiometry with the goal of studying a quantum phase transition from spin glass order to quantum paramagnetism.\cite{Wu1991,Wu1993}  This was achieved using linear ac susceptibility\cite{Wu1991} and also nonlinear susceptibility, $\chi_3$, measurements.\cite{Wu1993}  The results were, seemingly, an $(H_\perp,T)$ phase boundary with a classical phase transition at $T_g \simeq 130$ mK and a quantum phase transition at $H_C \simeq 1.2$ T.  Although at higher temperatures (lower fields), $\chi_3$ is found to exhibit a very sharp peak at the phase transition, at lower temperatures  (higher fields), the $\chi_3$ feature becomes extremely broadened.  Recent theoretical work\cite{Schechter2005,Tabei2006,Schechter2006,Tabei2008a,Schechter2008,Schechter2008b} has come to explain this effect through static random fields that are induced by the applied transverse field.  In fact, for any finite $H_\perp$, there is no longer a true spin glass state, but only domains of spin glass order that decrease in size with transverse field.  The phase boundary is, thus, a crossover rather than a true line of phase transitions.  However, for low fields where it is quite sharp, this distinction is not easily made experimentally.

At lower concentration, $x = 0.045$, Reich \emph{et al.}\cite{Reich1987,Reich1990} observed an unusual narrowing of the absorption spectra $\chi''(\omega)$ with decreasing temperature.  This is at odds with typical behavior of spin glasses\cite{Paulsen1987,Huser1983} and, in particular, with the clear broadening of $\chi''(\omega)$ that is seen at $x = 0.167$.\cite{Reich1990}    This stoichiometry has thus been referred to as an ``antiglass''.  In later work by Ghosh \emph{et al.}, the same research group, significantly more unusual results were observed.\cite{Ghosh2002,Ghosh2003}  In those results, the absorption spectra narrow appreciably at low temperatures and develop a strong asymmetry.\cite{Ghosh2002}  Hole-burning experiments were also performed where the system was saturated in the vicinity of a pump frequency, resulting in a hole in the spectrum as obtained with a variable probe frequency.\cite{Ghosh2002}  Most surprisingly, after cutting off the oscillating applied field, a ringing of the magnetization was observed, decaying over several seconds, reminiscent of coherent oscillations or a free induction decay.\cite{Ghosh2002}  Additionally, instead of a single broad feature in the specific heat, two sharp features were observed at around 100 mK and 300 mK and a very small percentage of the expected $R\ln 2$ entropy is accounted for with a numerical integral of $C/T$.\cite{Ghosh2003}  This unusual $C(T)$ and a $T^{-0.75}$ power law seen in the dc limit of $\chi$ were proposed to result from quantum entanglement of pairs of Ho$^{3+}$ moments.\cite{Ghosh2002}

In more recent years, the glassy portion of the phase diagram has sparked a great deal of controversy.  Recent specific heat measurements\cite{Quilliam2007} and ac susceptibility results\cite{Quilliam2008}, by the authors of this work, at $x = 0.045$, have not found any of the exotic antiglass features that were previously reported.  Instead, results consistent with spin glass physics were observed\cite{Quilliam2008}.  These results and others will be discussed in Section \ref{ResultsSection}  

J\"{o}nsson \emph{et al.}\cite{Jonsson2007} measured both an $x=0.045$ sample and an $x = 0.167$ sample, performing linear and nonlinear susceptibility measurements.  While not finding any major qualitative differences between those two samples, thus not observing a dramatic change to an antiglass state, they also argue that there is not sufficient evidence for a spin glass state, in either of the samples.  This finding opened up a debate regarding the size of magnetic field and sweep rate with which $\chi_3$ should be measured in order to make accurate conclusions regarding the existence of a spin-glass transition.\cite{AnconnaTorres2008,Jonsson2008}

From a theoretical point of view, it seems that the debate is largely settled.\cite{Tam2009,Gingras2011}  The Monte Carlo simulations that did not reveal a spin-glass transition\cite{Biltmo2007,Biltmo2008} focused on the Binder ratio $g$ as the mark  of spin glass freezing.  However, Tam and Gingras\cite{Tam2009} found that the correlation length $\xi_{SG}$ was a much better probe of a freezing transition as in conventional Edwards-Anderson spin glasses.~\cite{EArefs} Through a finite size scaling analysis, they showed a probable existence of a finite $T_g$ in a model largely representative of LiHo$_x$Y$_{1-x}$F$_4$.  Schechter and Stamp,\cite{Schechter2008b} meanwhile, have discussed the effects of off-diagonal components of the dipolar interaction and the nuclear hyperfine coupling and their introduction of quantum fluctuations, and have concluded that they are not sufficient to suppress spin glass ordering.

\section{Experiment}
\label{ExperimentSection}

High quality single crystals of LiHo$_x$Y$_{1-x}$F$_4$ were obtained commercially\cite{Tydex} and their characterization has been discussed previously.\cite{Quilliam2007}  Different parts of the same single crystals were used in Refs.~\onlinecite{Quilliam2007,Quilliam2008,Rodriguez2006,Rodriguez2010}.  Here, we have performed ac susceptibility measurements on $x = 0.018$, $x=0.045$ and $x=0.080$ samples, whereas we present specific heat measurements on four different samples: $x = 0.018$, 0.045, 0.080 and 0.12, the first three of which were measured previously in Ref.~\onlinecite{Quilliam2007}.

The ac susceptibility measurements described here were performed on a SQUID magnetometer, designed and implemented with glassy magnetic materials in mind, chosen for its flat response over many decades of frequency.  Where conventional susceptometers depend on inductive pick-up and the signal is proportional to frequency, a SQUID magnetometer, with a superconducting flux transformer, uses the Meissner effect which is frequency independent.  Additionally, the extremely high magnetic field sensitivity of a SQUID permits the use of excitation and pickup coils with very few turns, thereby avoiding resonances until frequencies well into the MHz range.  Since only small magnitude excitations need to be employed, significant heating within the sample can be fairly easily avoided.  A less than 20 mOe peak-to-peak oscillating magnetic field was employed in these measurements.

Our particular magnetometer design consists of a 375-turn NbTi primary coil and a second-order (5-7-5 turn) gradiometer as a pickup coil.  The gradiometer is made from Nb wire and is connected to a hand-wound Nb input coil to the SQUID via superconducting contacts, forming a flux transformer.  A NbTi trim coil (in parallel with the excitation coil) is included at one end of the gradiometer to adjust the balance of the circuit.  Coils are wrapped on phenolic coil forms which are press fit together concentrically.

Several layers of noise shielding are employed.  Pb superconducting shields surround the SQUID and the entire magnetometer, respectively.  The 1 K radiation shield of the cryostat is plated with Pb, and two $\mu$-metal shields, one in the He bath and one outside the dewar, are employed.  The $\mu$-metal shields also ensure that the sample and SQUID are cooled in extremely low fields (expected to be less than 5 mOe).  The SQUID is run in a flux locked loop with a 100 kHz modulation signal.\cite{ezsquid}  

Samples are heat sunk to a sapphire rod which is in turn heat sunk to the mixing chamber of a $^3$He/$^4$He dilution refrigerator.  Most of the susceptibility data presented here were taken on samples cut to be needle-shaped, with the Ising or $c$-axis along the length of the crystal.  Specifically, the needle-like sample geometries for the 1.8\%, 4.5\% and 8.0\% samples were $0.61\times0.89\times5.8$ mm$^3$, $0.57\times 0.77\times7.7$ mm$^3$ and $0.66\times 0.94\times 7.44$ mm$^3$, respectively.  Because of the large magnetic moments on the Ho$^{3+}$ sites, the demagnetization effect is very important to consider in these materials, particularly as $x$ gets larger.  Specifically, the demagnetization effect mixes the the in-phase and out-of-phase susceptibilities, resulting in an appreciable phase shift at lower temperatures and distorting the measured temperature dependence of the time constants of relaxation in the system.\cite{Dekker1989}  This makes it crucial to obtain a good calibration of the susceptibility and then correct for demagnetization.

For the 4.5\% and 8.0\% samples, calibration was obtained by comparing with different sample geometries ($0.57\times0.77\times3.3$ and $0.66\times0.94\times2.6$ mm$^3$ respectively).  An overall amplitude calibration pre-factor for each geometry was adjusted so that the results overlapped well after the demagnetization correction was applied.  For a more in-depth description of this method of demagnetization correction, see the appendix of Ref.~\onlinecite{Quilliam2011HTO}.  In the case of the 1.8\% sample, where the moments are sufficiently dilute that the demagnetization correction is not significant, calibration was performed by comparing to a superconducting Pb calibration standard.  A third geometry of the 4.5\% sample ($0.57\times0.77\times1.2$ mm$^3$) was also measured in order to thoroughly test for possible geometry effects, beyond those expected from the standard demagnetization effect.

Specific heat measurements, described in some detail previously~\cite{Quilliam2007}, were performed using the quasi-adiabatic heat pulse technique with a long time constant of equilibration.  A RuO$_2$ thermometer and metal-film heater are fixed directly to the samples, avoiding the use of a substrate and possible heat leaks.  Leads to the heater and thermometer are made from 6 $\mu$m diameter, superconducting NbTi filaments, chosen for their extremely low thermal conductance.  Thermal links from the dilution refrigerator to the sample are made from Manganin or PtW wire and the time constant of equilibration, $\tau_2$, is chosen to be greater than 1 hour at all temperatures.  The time constant of internal thermal relaxation, $\tau_1$, is always much shorter than the external time constant, $\tau_1$, thereby minimizing thermal gradients within the sample.  Twisted pairs, extensive heat sinking of leads and $\pi$-filters are employed to avoid self-heating in the thermometer.  The sample is suspended from thin nylon threads ($\sim 10$ $\mu$m diameter) inside a copper radiation shield that is heat sunk to the mixing chamber.  The addendum, due primarily to weak link and chip resistors, is determined to be $< 0.1$\% of the total heat capacity in any of the conditions presented here.  The thermometer resistance is measured with an \emph{LR700} ac resistance bridge.

\section{Results}
\label{ResultsSection}

\subsection{Magnetic Susceptibility}

\begin{figure}
\begin{center}
\includegraphics[width=3in,keepaspectratio=true]{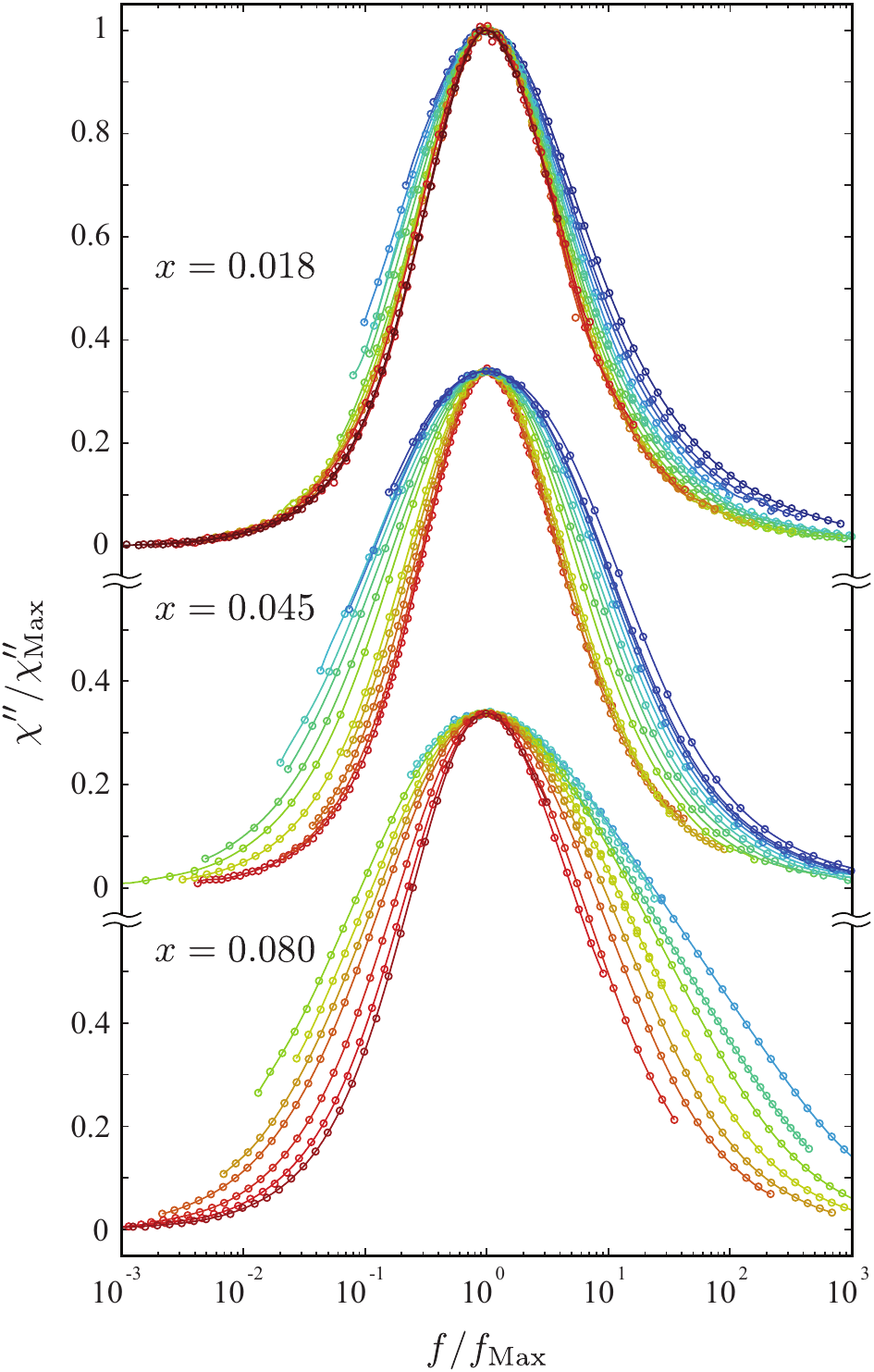}
\caption{(color online) Superimposed frequency scans of the absorption spectra (after correcting for the demagnetization effect) for the three different stoichiometries, obtained by normalizing $\chi''$ by the maximum, $\chi''_\mathrm{Max}$ and normalizing the frequency, $f$, by the peak position $f_\mathrm{Max}$.  The same legends of Fig.~\ref{FrequencyScans} apply here, with high temperatures as red and low temperatures as blue.  A clear broadening of the spectra is observed with decreasing temperature.  Also clear, is the increasing width of the spectra (in these temperature ranges) with increasing concentration, $x$.
\label{SuperimposedScans}
}
\end{center}
\end{figure}

Detailed frequency scans of the magnetic susceptibility, $\chi(f) = \chi'(f) - i\chi''(f)$, have been taken at various temperatures in between $\sim 80$ mK and $\sim 350$ mK, on three different stoichiometries at Ho concentrations of 1.8\%, 4.5\% and 8.0\%.  These scans, shown in Fig.~\ref{FrequencyScans}, exhibit qualitatively what is expected of glassy relaxation, in all three samples.  Broad (full width at half maximum ranging from 1.4 to 3 decades) and largely \emph{symmetric} peaks are seen in the absorption spectra, $\chi''(f)$, coinciding with an infection point in $\chi'(f)$.  The peak position, defined by $f_\mathrm{Max}$, drops sharply with decreasing temperature.  

The temperature ranges studied have been roughly chosen as those temperatures for which the frequency window of our measurement (from 1 mHz to around 1 kHz) permits the observation of the peak position in $\chi''$.  Surprisingly, the resulting temperature ranges for all three samples are quite similar.  This suggests an unusual lack of scaling of the relevant temperature ranges with concentration, which will later be explained as a result of dramatically changing time scales.

In order to compare the widths of the absorption spectra, they may be superimposed by plotting a normalized susceptibiltiy, $\chi''(f)/\chi''_\mathrm{Max}$, against frequency scaled by the peak frequency, $f/f_\mathrm{Max}$.  The resulting plots are shown in Fig.~\ref{SuperimposedScans}.  A clear, and largely symmetric, broadening of the spectra is seen with decreasing temperature.  It can also easily be observed that the spectra become, overall, wider as the concentration of Ho$^{3+}$ ions is increased from 1.8\% up to 8.0\%.  Certainly, for most temperatures, the spectrum of an $x=0.08$ sample will be broader than that of an $x=0.045$ sample, for example.  All three samples appear to tend toward a common shape and width of absorption spectrum at higher temperatures.

\begin{figure}
\begin{center}
\includegraphics[width=3.25in,keepaspectratio=true]{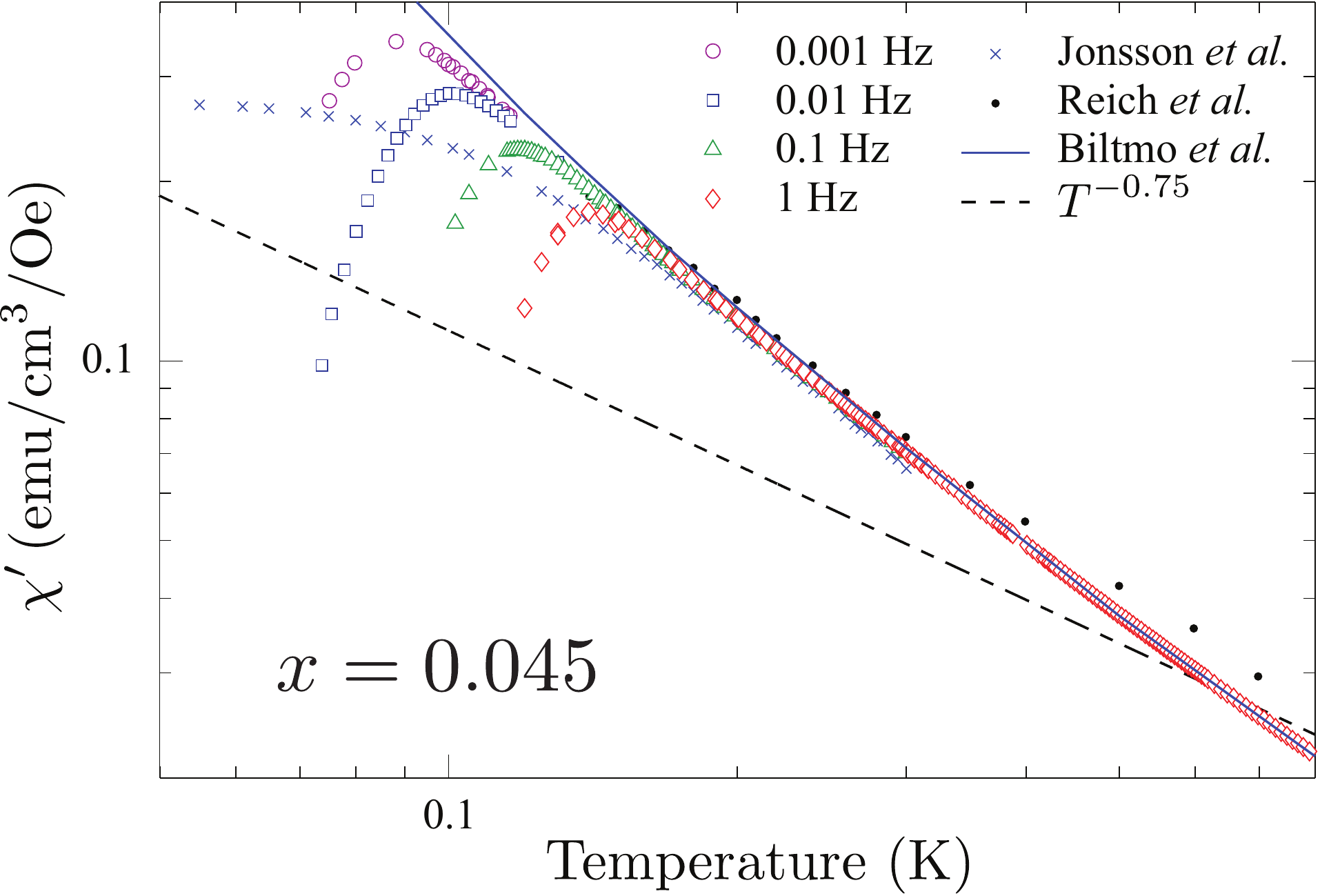}
\caption{(color online) Temperature scans of the demagnetization-corrected, in-phase susceptibility, $\chi'(T)$, of an $x = 0.045$ sample of LiHo$_x$Y$_{1-x}$F$_4$ at four different frequencies of measurement: 0.001 Hz (purple), 0.01 Hz (blue), 0.1 Hz (green) and 1 Hz (red).  Also shown, for comparison, are experimental data from J\"{o}nsson \emph{et al.}\cite{Jonsson2007} (which we have corrected for demagnetization assuming negligible $\chi''$), experimental data from Reich \emph{et al.}\cite{Reich1990}, numerical simulation data from Biltmo and Henelius\cite{Biltmo2007} (with an arbitrary scale factor) and a $T^{-0.75}$ power law proposed by Ghosh \emph{et al.}\cite{Ghosh2003}.
\label{TempScans}
}
\end{center}
\end{figure}

Dense temperature scans of the susceptibility, that have been performed only on the $x = 0.045$ sample, are shown in Fig.~\ref{TempScans}.  The in-phase susceptibility $\chi'$ is found to increase monotonically as the temperature is reduced until a maximum at the frequency-dependent freezing temperature $T_f(f)$, below which it drops out sharply.  $\chi'(T)$ is shown at four frequencies of measurement, with $T_f(f)$ showing a comparable behavior to the inverse of the function $f_\mathrm{Max}(T)$.  

A commonly employed parametrization of glassy relaxation, often referred to as the Mydosh parameter,\cite{Mydosh} is obtained through the ratio of the fractional change of freezing temperature to the logarithm of the change in measurement frequency, thus $\xi = \Delta T_f / (\tilde{T_f} \Delta \ln f)$ where $\tilde{T_f}$ represents either an average or limiting value of $T_f(f)$.  One could equally well apply this approach to the frequency scans of Fig.~\ref{FrequencyScans}, by changing the definition to $\xi = \Delta T/ (\tilde{T} \Delta \ln f_\mathrm{Max})$.  The resulting values, when applied to the very low end of our temperature (frequency) range, are 0.13, 0.11 and 0.11 for the 1.8\%, 4.5\% and 8.0\% samples, respectively.  Clearly the value of this parameter is dependent on the choice of temperature range employed, but serves as a rough order-of-magnitude estimate.  The canonical spin glasses tend to exhibit values of $\xi$ in the range of 0.005 to 0.06 and, as a rule of thumb, larger values have been taken as an indication of superparamagnetism or a lack of a finite-temperature glass transition, $T_g$.\cite{Mydosh}  This parametrization, if taken to be relevant, does not strongly support the idea that our samples of LiHo$_x$Y$_{1-x}$F$_4$ are spin glasses.  However, we argue that the parameter $\xi$ is not a measure of the existence of a finite $T_g$ but is, rather, a measure of how close to $T_g$ one is able to measure.  If a measurement were limited to frequencies in the MHz range, instead of the mHz range, even the canonical spin glasses would exhibit a large value of $\xi$ since relaxation with such high frequencies would occur very far away from $T_g$.  In other words, the Mydosh parameter relies heavily on the assumption that all systems have similar microscopic spin flip rates and that all measurements are performed on similar time scales.  We will show that this is not a safe assumption for the present system.

\subsubsection{Dynamical Scaling}

\begin{figure}
\begin{center}
\includegraphics[width=3.25in,keepaspectratio=true]{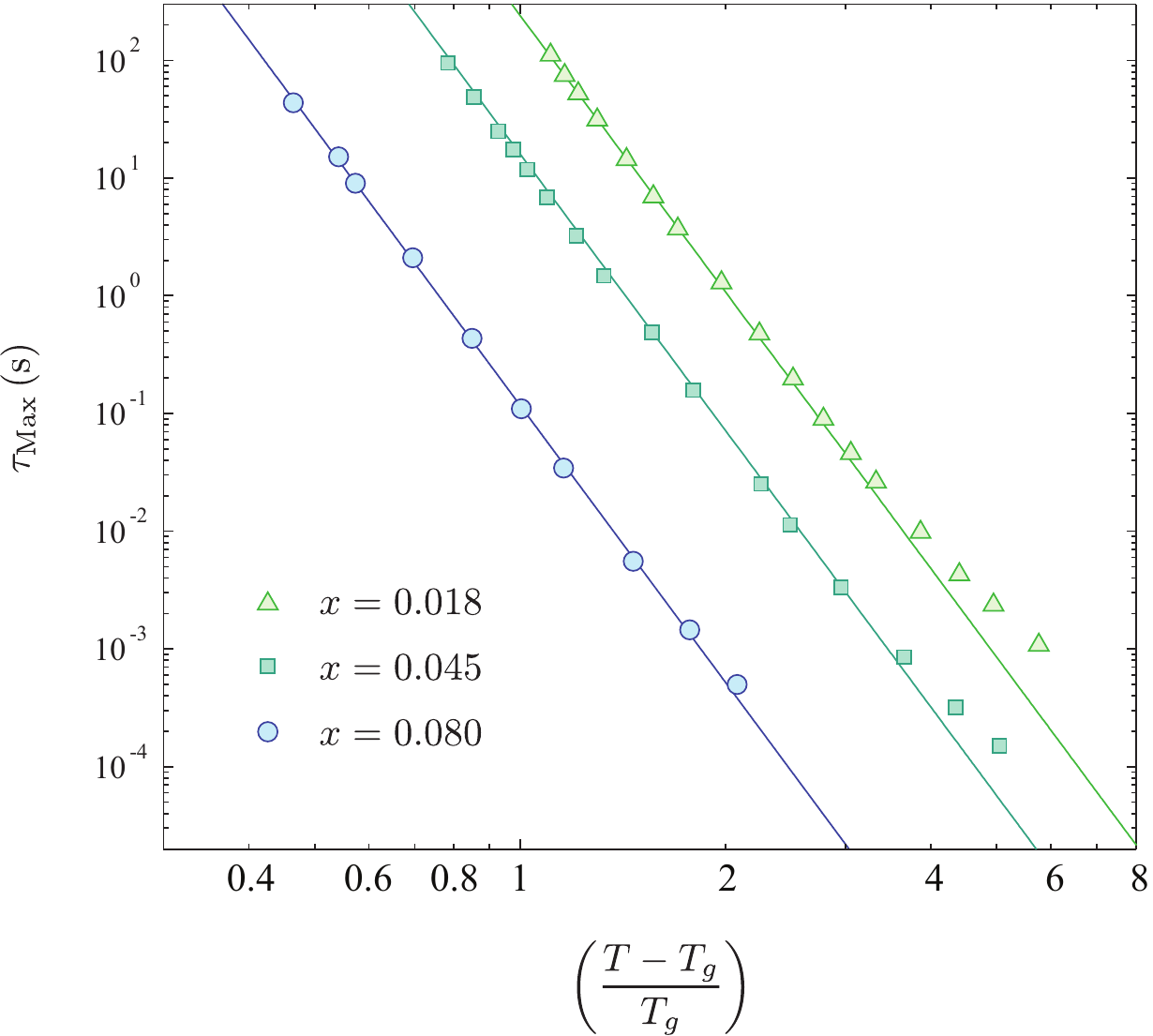}
\caption{(color online) Dynamical scaling plots showing that $\tau$ plotted against $(T-T_g)/T_g$ on a log-log scale can be well fit by a straight line.  The increase in $\tau_0$ with decreasing $x$ can be seen in the horizontal shift of the lines.  The fits obtained with no free parameters are shown for $x=0.045$ and $x=0.080$, where for $x = 0.018$, we have shown the fit obtained with fixed $z\nu = 8$.  
\label{DynamicalScaling}
}
\end{center}
\end{figure}

Perhaps a more quantitative way of determining whether samples of LiHo$_x$Y$_{1-x}$F$_4$ exhibit a spin-glass transition or not is to perform a dynamical scaling analysis where
\begin{equation}
\tau(T) = \tau_0 \left( \frac{T - T_g}{T_g} \right)^{-z\nu}.
\label{DynamicalScalingLaw}
\end{equation}
The choice of how to define $\tau(T)$, experimentally, is not entirely clear.  Perhaps the most rigorous approach would be to employ the limit $\lim_{\omega\rightarrow 0} \chi''(\omega)/\omega\chi'(\omega)$.\cite{Ogielski1985}  However, this limit is difficult to reach for most temperatures, generally requiring immensely low frequencies of measurement.\cite{PhDThesis}  Several research groups have also employed the maximum in $\chi'(T)$, or the freezing temperature $T_f$, as a function of frequency, in order to quantify the dynamics of a system.\cite{Maletta1982,Bontemps1984}  Defining a maximum in temperature is a slow process, however, and also restricts measurements to rather high temperatures in this particular system.  We have, therefore, chosen to parametrize the time scales of LiHo$_x$Y$_{1-x}$F$_4$ with the maximum of $\chi''(f)$, thus $f_\mathrm{Max}$, or equivalently, $\tau_\mathrm{Max} = 1/(2\pi f_\mathrm{Max}$).  These results are exhibited in Fig.~\ref{DynamicalScaling}.

Such a dynamical scaling analysis has already been presented for our $x = 0.045$ sample in Ref.~\onlinecite{Quilliam2008}.  Below 200 mK, a power law of the form of Eq.~(\ref{DynamicalScalingLaw}) was successfully fit to the data.  With no fixed parameters, the results of such a fit are $T_g = 42 \pm 2$ mK, $z\nu = 7.8 \pm 0.2$ and $\tau_0 = 16\pm 7$ s.  The critical exponent, $z\nu$, so obtained, matches extremely well with the value determined from Monte Carlo simulations of a short-range 3d Ising spin glass model~\cite{Ogielski1985} and also with values determined for several canonical spin glass systems.\cite{Bontemps1984,Bontemps1986,Vincent1986,Hamida1986}  This provides strong evidence that there is indeed a spin-glass transition in the system.  It is important to note that the intrinsic time constant of the system, $\tau_0 \simeq 16$ s, is exceptionally long compared to most other spin glass systems that have been studied.  For example, in Eu$_{0.4}$Sr$_{0.6}$S, a time constant of $\tau_0 \simeq 2\times 10^{-7}$ s is measured, thus it is about 8 orders of magnitude faster.\cite{Bontemps1984}  Such a $\tau_0$ as is observed here is largely unprecedented, although very recently a time constant of $\tau_0 \simeq 0.01$ s has been measured in another Ising spin glass Dy$_x$Y$_{1-x}$Ru$_2$Si$_2$.\cite{Tabata2010}  

This extraordinarily long time constant implies that, on reasonable time scales of an experiment, it is impossible to approach the glass transition and maintain equilibrium.  The lowest temperature data presented here, at 77 mK, are only at a reduced temperature $t = 0.8$.  Because of the very large exponent $z\nu$, in order to obtain equilibrium data at $t=0.1$ (10\% above $T_g$), for example, one would need to perform a measurement with a time constant of around 32 years.  This explains why the Mydosh parameter $\xi$ is larger here than in most spin glasses: quite simply we are far from the transition so the normalized slope of $\tau(T)$ is still relatively small.  Above 200 mK the dynamical scaling law no longer seems to hold.  However, since 200 mK corresponds to a reduced temperature $t\simeq 4$, this is perhaps not surprising and we are simply reaching the limits of critical behavior.  

In order to gain further insight into the reasons for such an anomalously long time constant, we have expanded our measurements to adjacent stoichiometries.  The $x = 0.080$ sample in fact lends itself equally well to such a treatment.  A successful dynamical scaling fit gives, again, a reasonable exponent of $z\nu = 7.8$.  The glass temperature obtained is $T_g  = 65\pm 3$ mK, showing the expected increase in $T_g$ with concentration $x$.  The intrinsic time constant, $\tau_0 \simeq 0.1$ s, is actually quite a bit smaller than that of the 4.5\% sample, thus there appears to be an inverse correlation of $\tau_0$ with $x$.

The 1.8\% sample does not present such a clear picture, however.  A fit with no free parameters yields $T_g = 41$ mK, $z\nu = 7.3$ and $\tau_0 = 61$ s.  The critical exponent is somewhat small and, most importantly, the glass transition, so obtained, barely differs from that of the 4.5\% sample, with more than twice the average interaction strength.  Fixing the critical exponent at $z\nu = 8.0$ also results in a perfectly adequate fit, with $T_g \simeq 35$ mK and $\tau_0 = 560$ s.  From a $\chi^2$ analysis, the fit is found to be quite under-constrained, with perfectly acceptable fits obtained with a range of glass temperatures from 33 to 46 mK.  This may be a result of only being able to measure far from the glass temperature.  

Even though the error bar on $T_g(x=0.018)$ is rather sizable, all the analyses above give surprisingly high glass temperatures.  If one were to assume that $T_g$ should scale linearly with $x$, we would expect $T_g \simeq 17$ mK.  A dependence not precisely linear, but not far from linear either, would be expected: some spin glass systems\cite{Mulder1981} scale roughly as $x^{0.7}$, the mean-field theory of LiHo$_x$Y$_{1-x}$F$_4$ has a slightly higher than linear dependence\cite{Stephen1981} and Monte Carlo simulations of Tam and Gingras\cite{Tam2009} suggest a roughly linear dependence on $x$.  We are therefore led to speculate that there is either some reason that $T_g$ is unusually high for the 1.8\% sample or that there is some additional complexity to the curve $\tau(T)$ which obscures the true spin glass temperature.   

What is very clear is that $\tau_0$ of the 1.8\% sample is much longer than that of the 4.5\% and 8.0\% samples, again highlighting the inverse correlation of $\tau_0$ with $x$.   In Section~\ref{DiscussionSection}, we discuss a possible explanation for this inverse behavior and the exceptionally slow dynamics in these materials.  

\subsubsection{Shape of Spectra} 

These materials clearly present challenging measurements, requiring very long time scales.  But with these slow dynamics comes an advantage in that one can resolve a larger part of the spectra with susceptibility experiments.  For most spin glasses, the frequency window of a typical ac susceptibility measurement corresponds to temperatures quite close to the glass temperature.  The resulting absorption spectrum, $\chi''(f)$, consists of a feature that is immensely broad, covering many decades.  One is able, for certain temperatures, to resolve the peak position but not say much more about the shape of that feature.\cite{Paulsen1987,Huser1983,Reich1990} Because the dynamics are inherently very slow in LiHo$_x$Y$_{1-x}$F$_4$, ac susceptibility measurements have a frequency window that corresponds to temperatures further away from $T_g$, where the absorption spectrum is quite a bit narrower.  This permits us to describe the shape of the spectra in more detail, including the widths and the power-law behavior of the low- and high-frequency tails.

\begin{figure}
\begin{center}
\includegraphics[width=3.25in,keepaspectratio=true]{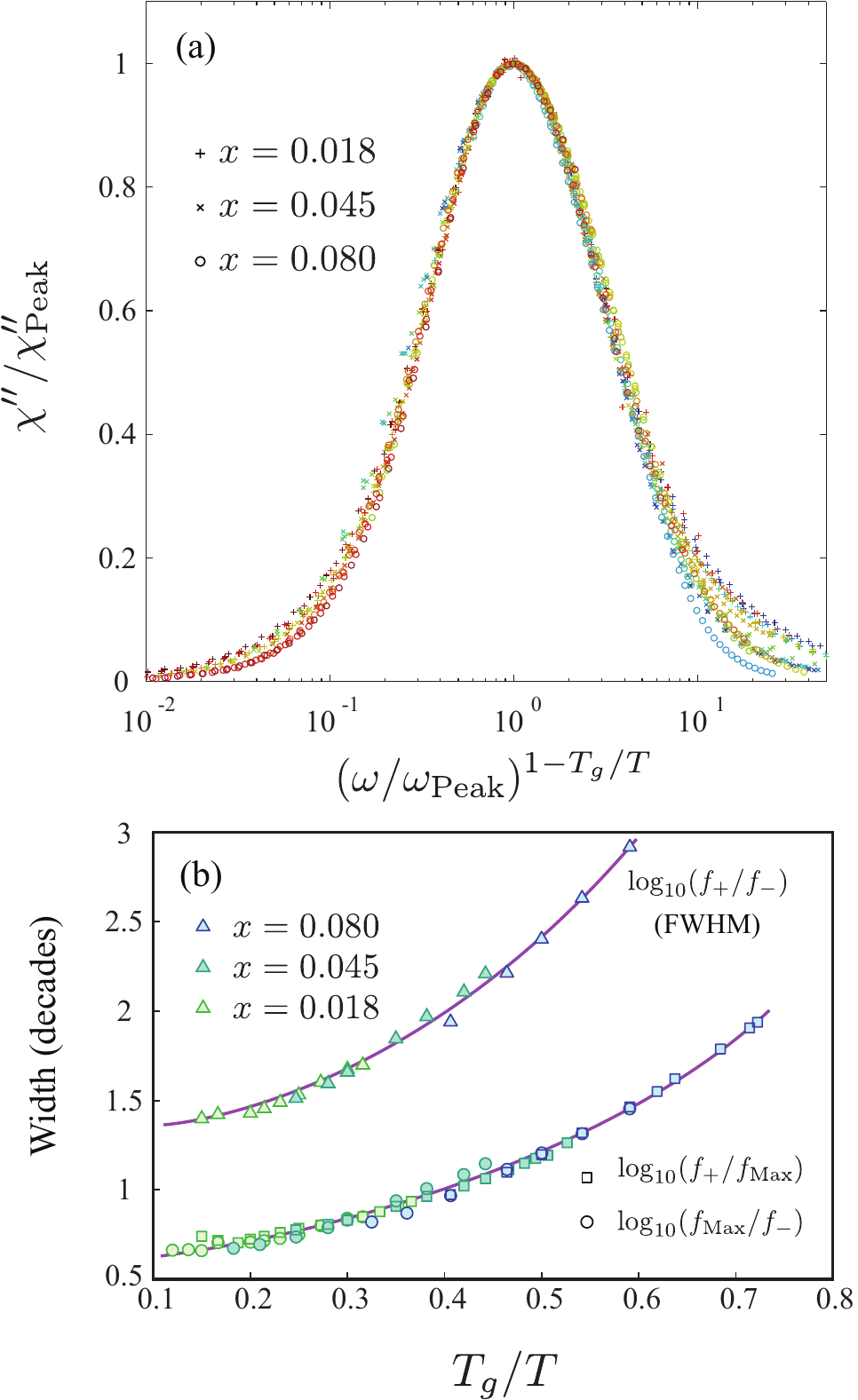}
\caption{(color online) The normalized absorption spectra of all three samples studied, when scaled on the frequency axis with the exponent $1-T_g/T$ are found to collapse onto a single curve, for much of the frequency range.  The tails of the absorption spectra are not accommodated with this scaling relation.  The same values of $T_g$ as were found in a dynamical scaling analysis are used to scale the 4.5\% and 8.0\% samples, whereas the 1.8\% sample required a $T_g = 30$ mK to obtain good overlap.  Colors of the data points correspond to those of Fig.~\ref{FrequencyScans} and range from $\sim 200$ mK (red) down to $\sim 80$ mK (blue).  All data have been corrected for the demagnetization effect.
\label{ScalingPlot}
}
\end{center}
\end{figure}

Most notably, all of the spectra, on all three samples, appear to tend toward a common form in the high temperature limit.  This form has a full width at half-maximum (FWHM) of roughly 1.4 decades.  The low-frequency tail appears to be linear in frequency.  In contrast, the high-frequency tail seems to follow a shallower power law $\propto f^{-\beta}$ with $\beta$ roughly 0.7 to 0.75.  As the temperature is lowered, the high-frequency power law becomes less steep.  At the lower end of our temperature range, the 4.5\% sample, for example, exhibits a high-frequency tail with an exponent closer to $\beta \simeq 0.65$.    Although the 8.0\% sample broadens much more at the temperatures studied here, the high-frequency tail does not get much shallower.  Meanwhile, the 1.8\% sample shows very little overall broadening as the temperature is reduced, yet has a high-frequency tail with an exponent close to $\beta\simeq 0.55$.   It is difficult to determine whether the low-frequency tail also becomes less steep or whether it is simply that the spectra become very broad and we are unable to observe the limiting behavior on that side.

These details make it difficult to fit successfully to the spectra.  The Davidson-Cole form,\cite{Davidson1950,Davidson1951} for example, can reproduce the low and high-frequency power-law behavior, but cannot account for the widths of the spectra, although it comes quite close at higher temperatures.  The ac susceptibility resulting from relaxation processes can be described by the Debye model
\begin{equation}
\chi(\omega) = \chi_0 \int_{-\infty}^\infty \frac{\rho(\tau)d\ln\tau}{1-i\omega\tau}
\end{equation}
where $\rho$ is a distribution of relaxation times of the system.  An absorption spectrum that is linear at low frequency, as is found in the Davidson-Cole form, results from a cut-off at long $\tau$ (low frequency) in $\rho(\tau)$.  The distribution $\rho$ must also have a long low $\tau$ (high frequency) tail with power law $\beta$ in order to give rise to that same exponent in the susceptibility at high frequencies.\cite{Matsuhira2001}

In looking at Fig.~\ref{SuperimposedScans}, it can be noticed that the width of the spectra (on a log scale) is, at least roughly, correlated with the distance from the transition at $T_g$.  Since the spectrum widths appear to tend toward a constant value at high temperatures, we make the ansatz that the excess width at lower temperatures is simply inversely proportional to the ratio $T/T_g$.  This can be tested by plotting $\chi''/\chi''_\mathrm{Max}$ against $ (f/f_\mathrm{Max})^{(1-T_g/T)}$ which is shown in Fig.~\ref{ScalingPlot}(a).  For the 4.5\% and 8.0\% samples, a good collapse of the data over a large range of frequencies is obtained, using the same values of $T_g$ as were obtained from the dynamical scaling analysis.  In fact, those values of $T_g$ seem to optimize the overlap of the data, lending further evidence that they are indeed the correct transition temperatures.  The 1.8\% sample is again less straightforward.  A choice of $T_g = 30$ mK, allows a good overlap.  

Alternatively, as shown in Fig.~\ref{ScalingPlot}(b), we can plot the full widths or half widths (at half the maximum) of the spectra in log of frequency.  Here we define the maximum frequency as $f_\mathrm{Max}$ and the left and right half maxima as occurring at $f_-$ and $f_+$ respectively.  Thus the full width at half the maximum (FWHM) is given by $\log_{10}(f_+/f_-)$.  The half widths can be given by  $\log_{10}(f_+/f_\mathrm{Max})$ and  $\log_{10}(f_\mathrm{Max}/f_-)$.  Plotting all of these quantities as a function of $T_g/T$ shows that the widths of the spectra do indeed follow a very simple behavior.  While quite successful for the FWHM and much of the peak in $\chi''$, this scaling does not permit good agreement in the low and especially the high-frequency tails of the spectra.  This is again illustrative of the unusual, shallow tails of the spectra.

\subsection{Specific Heat}

It has been understood for some time\cite{Mennenga1984} that the low-temperature specific heat of LiHo$_x$Y$_{1-x}$F$_4$ is dominated by the nuclear contribution, the result of a strong nuclear dipole hyperfine coupling to the electronic moments.  Note, the nuclear electric quadrupolar coupling is relatively insignificant~\cite{Abragam}.  The resulting Schottky-like feature is quite large as a result of the number of degrees of freedom afforded by the $I=7/2$ nuclear moment.  If the electronic moments are considered to be perfectly classical or Ising-like, the moments can be written simply as the spin-1/2 variables $S_i^z$.  There are then no off-diagonal components of the hyperfine interaction and its contribution to the Hamiltonian will be 
\begin{equation}
\mathcal{H}_{HF} = \sum_i A_\| m_i S^z_i
\end{equation}
where $A_\| = 2 \langle J_z \rangle A$.  The summations over the $m_i$ variables in the partition function, $\mathcal{Z}$, can simply be rewritten as summations over $x_i = m_iS^z_i$ in order to write $\mathcal{Z}$ as the product of an electronic part $\mathcal{Z}_0$ and an electronuclear or hyperfine part, $\mathcal{Z}_{HF}$, defined as
\begin{equation} \mathcal{Z}_{HF} =  N\sum_{x = -7/2}^{7/2} e^{A_\| x /2k_B T}.  \end{equation}
Then, since all thermodynamic quantities are derived from $\ln\mathcal{Z} = \ln\mathcal{Z}_0 + \ln\mathcal{Z}_{HF}$, we can, for instance, take the specific heat to be a summation of a nuclear hyperfine or single-ion contribution $C_{SI}$ and a contribution from only the electronic magnetic moments, $C_m$.  

However, the situation described above is an oversimplification of the problem.  In fact, transverse components of the nuclear hyperfine interaction do come into play if one performs a diagonalization of the full single-ion Hamiltonian including crystal field, hyperfine and Zeeman Hamiltonians, $\mathcal{H}_{CF} + \mathcal{H}_{HF} + \mathcal{H}_Z$, a $136 \times 136$ matrix (8 nuclear energy levels $\times$ 17 crystal field levels).  The form for $C_{SI}$ used here has been obtained by performing such a diagonalization.  That said, with transverse hyperfine effects included, the partition function cannot be factored and a subtraction of the single-ion specific heat is no longer fully justified.  Nonetheless, we will continue to present data after such a subtraction in order to compare with other experimental works and to obtain an approximate measure of the system in the absence of nuclear moments.

\begin{figure}
\begin{center}
\includegraphics[width=3in,keepaspectratio=true]{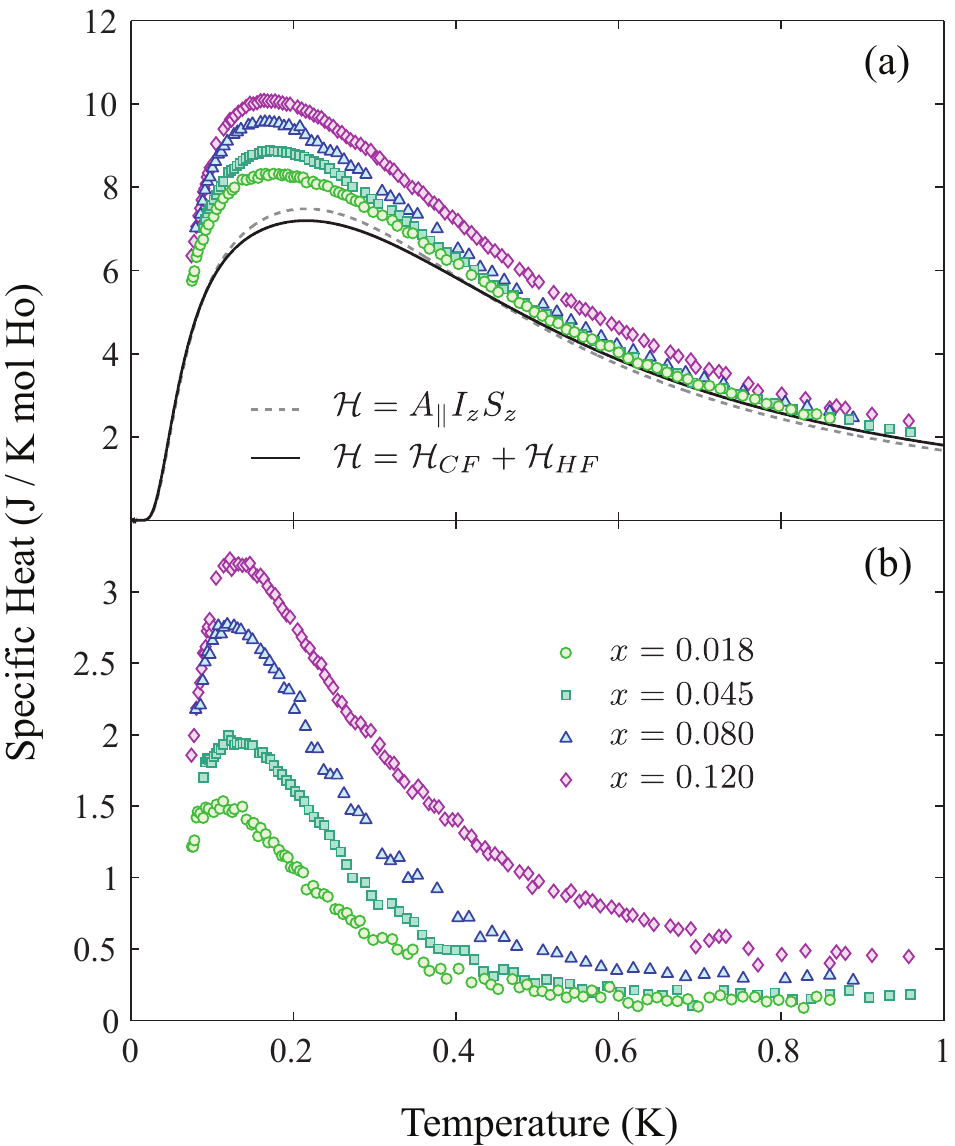}
\caption{(color online) (a) Total specific heat measured on four samples of LiHo$_x$Y$_{1-x}$F$_4$ at $x = 0.018$, 0.045, 0.080 and 0.12.  Also shown are the calculated single ion contributions.  The dashed line assumes an Ising doublet with no crystal field excitations.  The solid curve is a diagonalization of the crystal field and hyperfine Hamiltonians, thus includes some coupling with crystal field levels induced by the transverse components of $\mathcal{H}_{HF}$.  The latter is subtracted from the total specific heat to yield the specific heat of the magnetic moments, $C_m(T)$, which is plotted in (b).
\label{SpecificHeat}
}
\end{center}
\end{figure}

The measured, total specific heat of four different stoichiometries of LiHo$_x$Y$_{1-x}$F$_4$ is shown in Fig.~\ref{SpecificHeat}(a).  The specific heat is, as expected, clearly dominated by nuclear effects.  The single-ion specific heat, $C_{SI}$, is shown as the black curve.  Performing a subtraction of $C_{SI}$ leaves us with broad peaks, as seen in Fig.~\ref{SpecificHeat}(b).  A broad peak in the specific heat is expected in spin glass materials as the corresponding critical exponent $\alpha$ is generally quite negative, in the range $-2$ to $-4$.\cite{Ogielski1985,FischerHertz}  However, one does expect the maximum to be close to, often 20\% above $T_g$.  This is clearly not the case here, with the peak position at $\sim 120$ mK in all samples, independent of $x$ and therefore independent of $T_g$.  

It can also easily be noticed that the magnetic specific heat $C_m$ is overall much smaller for lower concentration samples.  Since the entropy may be obtained from the relation $S = \int_0^T (C/T)dT$, there is obviously much less entropy accounted for in the $x = 0.018$ sample than in the $x = 0.12$ sample, for example.  This suggests that there is significant residual entropy at the lowest temperatures of our measurement.  What happens below that temperature ($\sim 75$ mK) decides what the true ground state, residual entropy $S_0$ might be.  Almost 100\% of the expected $R\ln 2$ entropy is reproduced in the $x = 0.12$ sample, and a steeper than linear extrapolation to $T=0$ (common in spin glasses~\cite{Mydosh}) is required in order to not exceed $R\ln 2$.  However, the other samples do seem as if they might possess varying levels of $T=0$ residual entropy.  

\section{Discussion} \label{DiscussionSection}

\begin{figure*}
\begin{center}
\includegraphics[width=7in,keepaspectratio=true]{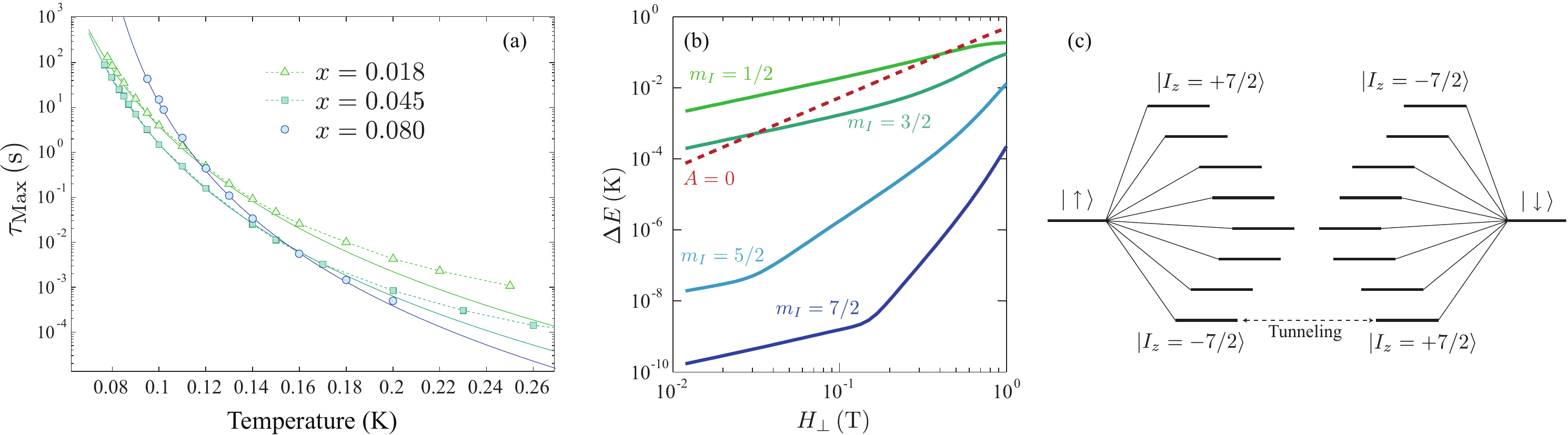}
\caption{(color online) (a) The time constants $\tau_\mathrm{Max}$ as a function of temperature for the three samples of LiHo$_x$Y$_{1-x}$F$_4$ studied here.  The solid lines are the dynamic scaling fits, described in the text, whereas the dotted lines are simply guides to the eye.  It can be noticed that the time constant is inversely proportional to concentration $x$ at temperatures above 150 mK.  Below that point, the curves begin to cross as time scales in higher $x$ samples begin to diverge with a power-law behavior at higher temperatures.  (b) The energy splittings versus transverse field of the four lowest lying time-reversed pairs of electronuclear states, obtained by diagonalizing the single ion Hamiltonian $\mathcal{H}_{SI} = \mathcal{H}_{CF} + \mathcal{H}_{HF} + \mathcal{H}_Z$ (solid lines).  The dashed line is the energy splitting in the ground-state doublet with no nuclear hyperfine interaction.  (c) An energy level diagram showing the 16 electronuclear states.  The horizontal distance between pairs is qualitatively representative of the strength of the tunneling matrix elements.   At low temperatures, all but the lowest two electronuclear states are depopulated and relaxation is a result of tunneling between these states.
\label{DiscussionFigure}
}
\end{center}
\end{figure*}

While the dynamical measurements presented here have provided appreciable evidence of spin glass physics, several puzzles remain to be solved.  Of utmost importance is an explanation for the incredibly slow dynamics that are observed well above the glass transition temperature $T_g$.   The specific heat is also rather confusing because of the stationary peak position as $x$ is changed.  Finally, the $T_g$ obtained here for the 1.8\% sample seems anomalously high when compared to values for the other stoichiometries and even the Curie temperature, $xT_C(x=1)$.  

We first attempt to understand the long time scales in this system, which are unprecedented in spin glass materials.  A possible explanation may come from a link to so-called superspin glasses.\cite{Djurberg1997}  In such materials, consisting of randomly interacting magnetic particles, dynamics have been observed that are noticeably slower than those of canonical spin glasses.\cite{Djurberg1997,Chen2005}  Since the magnetic units in a superspin glass system are often much larger (for example\cite{Sasaki2005} $\sim 300\mu_B$) than the single-spin units of a spin glass, the microscopic spin flip time can be much longer.  On the other hand, some objects defined as superspin glasses maintain rather fast spin flip times and values of $\tau_0$ comparable to those of canonical spin glasses.\cite{Suzuki2009}  Other systems of random magnetic particles, known as superparamagnets, show slow relaxation due to large anisotropy barriers, but not necessarily a finite-temperature freezing transition as in spin glasses.~\cite{Mydosh}  One could imagine that in LiHo$_x$Y$_{1-x}$F$_4$, there might be large ferromagnetically correlated regions that act as magnetic particles or clusters, resulting in superspin glass or superparamagnetic behavior.  However, one should expect such an effect to get stronger as one nears the ferromagnetic phase at higher $x$; here we see a marked speeding up of the dynamics as $x$ is increased.  Furthermore, the moments are evenly distributed throughout the sample in LiHo$_x$Y$_{1-x}$F$_4$ and the samples are extremely high quality single crystals with no indications of clustering.\cite{Quilliam2007}  Most likely, these material should be treated as spin glasses rather than superspin glasses, with single Ho$^{3+}$ moments as the fundamental building blocks. 

A probable explanation for the long time scales involved can be found in single ion physics.  Most importantly, the Ho$^{3+}$ ions in LiHo$_x$Y$_{1-x}$F$_4$, when considered individually, have a truly Ising doublet ground state wherein the matrix elements $\langle \uparrow | J_+ | \downarrow \rangle$ and $\langle \uparrow | J_- | \downarrow\rangle$ vanish.  The result is that spin flips are forbidden transitions and must occur via the next excited state at $\sim 11$~K, or must occur via quantum tunneling.  In the temperature range of our measurements, at 200 mK and below, there is very little probability of exciting an ion into the $|\gamma\rangle$ state, thus we are left with almost entirely tunneling, which has the potential to be quite slow.  A similar argument might apply to the Ising material Dy$_x$Y$_{1-x}$Ru$_2$Si$_2$ which shows a similarly long $\tau_0$ of 0.01 s.\cite{Tabata2010}

In the absence of off-diagonal contributions to the Hamiltonian, this tunneling would not occur at all.  However, in a real material there are always mechanisms for spin relaxation.  The tunneling here is primarily governed by the off-diagonal or non-secular components of the dipolar interaction~\cite{Atsarkin1988,Chin2007}
\begin{equation}
\mathcal{H}_D^{ns} = \sum_{i\neq j} \left(   c_{ij} J_i^{x'}  J_j^z + d_{ij} J_i^{x'} J_j^{x'} \right)
\label{nonsecular}
\end{equation}
where the coefficients $c_{ij}$ and $d_{ij}$ are easily obtained from Eq.~(\ref{HamD}) and $x'$ is the projection of $\vec{r}_{ij}$ in the $xy$-plane.  Essentially, there is another criterion for strong Ising character of a material, that is that the dipolar interaction does not introduce a large degree of mixing with the next excited crystal-field state and therefore does not introduce a large degree of quantum tunneling.  This may be reduced to the requirement that $\langle \uparrow | \mathcal{H}_D | \gamma\rangle \ll 1$.  This requirement is satisfied in LiHo$_x$Y$_{1-x}$F$_4$ by virtue of the form of the states $|\uparrow\rangle$, $|\downarrow\rangle$ and $|\gamma\rangle$.  The pyrochlore system Tb$_2$Ti$_2$O$_7$, is an important counterexample.  Despite an energy gap $\Delta$ at least as large as that of LiHo$_x$Y$_{1-x}$F$_4$, the coupling with those excited states through $\mathcal{H}_D$ is much more significant, introducing quantum fluctuations and possibly making Tb$_2$Ti$_2$O$_7$ an interesting \emph{quantum spin ice} material.~\cite{Molavian2007}

Even once assured that this system is strongly Ising, there is some tunneling that allows spin flips to occur and gives rise to dynamics at low temperature.  Here, we engage in some basic calculations with the goal of understanding whether the immensely slow dynamics observed in dilute LiHo$_x$Y$_{1-x}$F$_4$ can be realistically attributed to largely single-ion physics.  We concentrate here on the first and dominant $J_i^{x'} J_j^z$ term of Eq.~(\ref{nonsecular}).  In order to greatly reduce the difficulty of the problem, we analyze a single Ho$^{3+}$ ion and incorporate that dominant off-diagonal term of $\mathcal{H}_D$ as a transverse magnetic field.  At any given time, nearby spins may be thought of as generating a random and fluctuating transverse field.  We approximate the level of this transverse field in two different ways.  (1) We simply take the field generated by a nearest neighbor Ho$^{3+}$ ion, $h_{nn}^\perp$.  (2) We take random distributions of ion positions and spin orientations, in other words snapshots of the system in time, calculate the transverse field at site $i$ as
\begin{equation}
\vec{h}_i^\perp \propto \sum_j \hat{x}' c_{ij} J_j^z
\end{equation}
and average $|h_i^\perp|$ over many different random configurations to obtain an average transverse field $\tilde{h}^\perp$.

While the importance of random fields in LiHo$_x$Y$_{1-x}$F$_4$ has been discussed extensively,~\cite{Tabei2006,Schechter2005} it has generally been \emph{static} random fields induced by an applied external transverse field $H_\perp$.  Since we are discussing the system in zero transverse field, above the spin-glass transition, the random fields are fluctuating in time along with the spins.  As such, they may not strongly affect the equilibrium properties of the material, but they certainly should have a powerful effect on the dynamics of the system.  

Essentially, the tunneling rate of an ion may be related to the energy splitting $\Delta E$ of the Ising doublet induced by the transverse field.~\cite{Schechter2008b}   Specifically, we can say that the time constant for a single spin flip can be approximated as $\tau_{SF} = \hbar/\Delta E$.  Since the dipolar interactions in a lower concentration ($x$) sample are on average lower in energy, so are the random transverse fields seen by the magnetic ions.  Thus we can immediately see why there is an inverse correlation between $x$ and $\tau$, as seen at the higher temperatures in Fig.~\ref{DiscussionFigure}(a).  Taking, for instance, the $\tau(T)$ curves for the 1.8\% sample and the 8\% sample, it can easily be seen that $\tau$ is much larger for the 1.8\% sample at higher temperatures.  The glass temperature $T_g$, however, is \emph{not} inversely correlated with $x$.  Thus as the temperature is lowered and $T_g(x=0.08)$ is approached, $\tau$ of the 8\% sample begins to diverge and there is a crossing of the curves.

Qualitatively, the above description seems rather plausible, but is it quantitatively sufficient to explain the magnetic properties of LiHo$_x$Y$_{1-x}$F$_4$?  First we consider the single-ion Hamiltonian, ignoring the nuclear hyperfine interaction, thus keeping the crystal field and transverse field Zeeman interaction: $\mathcal{H}_{CF} + \mathcal{H}_Z$.  We then exactly diagonalize this $17\times17$ Hamiltonian and determine the energy splitting between the two lowest lying states as a function of transverse field, shown as the red dotted line in Fig.~\ref{DiscussionFigure}(b).  The transverse field induced by a nearest neighbor spin has magnitude $h_\mathrm{nn}^\perp = 0.057$ T, thus leading to an energy splitting of 1.8 mK and a microscopic tunneling time of $\tau_{SF} = 4.2$ ns.  Taking the average transverse magnetic field generated by a random occupancy and spin configuration, gives $\tilde{h}^\perp = 0.026$, 0.052 and 0.081 T for $x = 0.018$, 0.045 and 0.08 respectively.  It seems clear that the tunneling rates determined in this way are far too fast to explain these materials' behavior.

To go further, one must introduce the energetically significant nuclear hyperfine interaction.  In several portions of the phase diagram of LiHo$_x$Y$_{1-x}$F$_4$, it has already been observed or theoretically explained that the nuclear hyperfine interaction blocks the effect of an applied transverse field at low temperatures.\cite{Bitko1996,Schechter2008b}  Also shown in Figure~\ref{DiscussionFigure}(b), are calculations performed by diagonalizing the full single-ion Hamiltonian in a transverse field, thus $\mathcal{H}_{CF} + \mathcal{H}_{HF} + \mathcal{H}_Z$, a $136\times 136$ matrix.  The nuclear hyperfine interaction splits each of the Ising states into eight	 electronuclear energy levels, separated by roughly 200 mK, as shown in Figure~\ref{DiscussionFigure}(c).  In the absence of magnetic field, there are degenerate pairs of time-reversed electronuclear states, the lowest of which consists of $|\uparrow,-7/2\rangle$ and $|\downarrow, +7/2\rangle$.  These pairs are split in energy by a transverse magnetic field, as shown for the four lowest-lying pairs in Fig.~\ref{DiscussionFigure}(b).  When compared to the calculation in the absence of $\mathcal{H}_{HF}$, one can see that the lowest energy pair experience a drastically smaller energy splitting, $\Delta E$, hence dramatically slower tunneling.  Certainly transitions between higher energy electronuclear states are either as fast or faster than spin flips without the nuclear hyperfine coupling, but these states will be very heavily depopulated at temperatures below 200 mK or so.   

In a perturbative approach, one can describe the dynamics in terms of virtual transitions.~\cite{Chin2007}  However, there is certainly no energy splitting between $|\uparrow,-7/2\rangle$ and $|\downarrow, +7/2\rangle$ to first-order in perturbation theory.  Such a splitting requires many orders in perturbation theory as one climbs up one ladder of nuclear energy levels and back down on the other side of the diagram in Fig.~\ref{DiscussionFigure}(c).  This is the reason that transitions are so immensely slow between the lowest-lying states and the reason the nuclear hyperfine interaction so effectively blocks the effects of transverse field.~\cite{Schechter2008b}

This importance of single-ion physics was first suggested by Atsarkin\cite{Atsarkin1988} who proposed that the slow relaxation seen in LiHo$_x$Y$_{1-x}$F$_4$ could be largely explained through depopulation of nuclear energy levels.  Further evidence of such a link comes from work of Giraud \emph{et al.}\cite{Giraud2001,Giraud2002,Giraud2003} who have studied samples that are more dilute than those studied here by more than an order of magnitude.  In those samples, for temperatures above 100 mK, one is primarily observing single-ion properties with very little impact of the surrounding ions and therefore relatively minor effects of disorder.  In such an experimental situation one is able to observe resonant tunneling of the electronic spins where nuclear energy levels are coincident with the Zeeman energy $g_J\mu_B\mathbf{H}\cdot \mathbf{J}$.\cite{Giraud2001,Giraud2002,Giraud2003}  Where the system is off-resonance, the dynamics become very slow.  Even at temperatures as high as 1 K significant absorption occurs with frequencies as low as 1 Hz.  Clearly the picture becomes progressively more complicated as one introduces larger and very random interaction strengths.

The behavior of the real system ought to be somewhat challenging to model accurately.  There are transitions between many different states to consider, longitudinal fields will also play a role, cotunneling of spins may be important and eventually the system must be thermally linked to a bath of phonons.  However, this simple single-ion picture can hopefully give an estimate of what the tunneling rates will be at low temperatures.  Taking the average internal transverse fields quoted above, we obtain estimated tunneling times of $\tau_{SF} = 20$, 9.8 and 6.1 ms for $x = 0.018$, 0.045 and 0.08 respectively.  This more than 6 order of magnitude increase in time constant as a result of the hyperfine interaction \emph{must} be the explanation for the exceedingly long time constants in this system.   

There remains, however, an appreciable discrepancy between these very simply calculated values and the measured time constants, $\tau_0$.  First we note that there is no particular reason that $\tau_0$ obtained from a dynamical scaling analysis should be representative of a single tunneling event.  If that were the case, one would have a $\tau$ much faster than the microscopic tunneling time as soon as the temperature was brought above $T = 2T_g$, which is quite unphysical.  Furthermore, it is evident that we have not yet reached the regime below which there is no temperature dependence to the tunneling rate.  This is clear from the fact that $\tau_0$ is exponentially dependent on $x$.  In other words, the microscopic tunneling rate is following an exponential Arrhenius law in temperature and since the critical region of a lower-$x$ sample is at lower temperatures, $\tau_0$ \emph{is exponentially dependent on} $x$.  Such estimates and qualitative conclusions would be greatly enhanced with precise theoretical work relating microscopic spin flip times with macroscopic critical behavior, calculations which have, to our knowledge, not yet been performed.  Regardless of the exact quantitative result, it is clear that the nuclear hyperfine interaction, at the lowest temperatures, leads directly to a slowing of the dynamics by many orders of magnitude. 

While unprecedented in spin glasses, we note that such extreme tunneling times have been observed in the spin ice material Dy$_2$Ti$_2$O$_7$ where the microscopic tunneling rate is found to be several ms$^{-1}$ even at 5 K or higher, even without a relevant nuclear hyperfine coupling.\cite{Jaubert2009,Snyder2004} Ho$_2$Ti$_2$O$_7$ has similarly long relaxation times.\cite{Quilliam2011HTO}  The primary reason that spin ice materials are so slow is that the next excited crystal field energy levels are at close to 300 K (as opposed to 10 K in LiHo$_x$Y$_{1-x}$F$_4$).  Since those materials are not randomly diluted, the internal transverse fields may at times be canceled by symmetry leading to even slower dynamics.

Two main puzzles remain in our LiHo$_x$Y$_{1-x}$F$_4$ results.  (1) The 1.8\% sample has an anomalously high $T_g$ (or apparent $T_g$), as compared to the other samples studied.  Perhaps the actual $T_g$ is really enhanced at low $x$ for some currently unknown reason.  Alternatively, $T_g$ may simply appear higher.  For the dynamical scaling picture to hold, the $\tau_{SF}$ must be somewhat flat in temperature.  Measurements on the 1.8\% sample may be in a regime where there is still a temperature-dependent population of higher nuclear energy levels.  The purely single-ion physics is already quite rich as a result of the nuclear hyperfine interaction and should consist of several Arrhenius laws before interactions take over at low $T$ and power-law behavior begins to dominate.  This rich nuclear contribution to the dynamics may complicate the interpretation of the $\tau_\mathrm{max}(T)$ curve and lead to an inaccurate determination of $T_g$.

(2) The other main puzzle remaining is the peak position of the specific heat curves and its lack of scaling with $x$, which contrasts with numerical treatments of the problem.~\cite{Biltmo2008}  This again may be connected to single-ion physics and the very important coupling to the nuclear moments.  It is certainly likely that the subtraction of $C_{SI}$ from the total $C$ does not reveal the behavior that would be expected without the hyperfine coupling.  At such low temperatures, the system involves electronuclear states that are not separable.\cite{Schechter2008b}  The peak seen at $\sim 120$ mK in all the samples is likely representative of that coupling.  Perhaps at much lower temperatures, closer to $T_g$, another peak could be found in $C$.  However, as with susceptibility, specific heat measurements will eventually also fall out of equilibrium thus performing measurements at lower $T$ than presented here is quite challenging and potentially misleading.

\section{Comparison with Other Groups}	

\subsection{Antiglass}

In work of Reich \emph{et al.}\cite{Reich1987,Reich1990} and later, by the same research group, Ghosh \emph{et al.},\cite{Ghosh2002,Ghosh2003} a highly unusual state was observed at a concentration of $x=0.045$, which has been referred to as the ``antiglass'' state, as a result of certain properties that are quite the opposite of those expected for a spin glass.  The phenomenology of the antiglass phase includes a narrowing and markedly increasing asymmetry of the absorption spectra as the temperature is reduced, a $T^{-0.75}$ power law of the dc limit of the susceptibility and sharp peaks in the specific heat.  In contrast, we are unable to reproduce the above effects, and find more conventional behavior that is consistent with a spin glass, for $x=0.045$ and two concentrations surrounding it.  

\begin{figure}
\begin{center}
\includegraphics[width=3.25in,keepaspectratio=true]{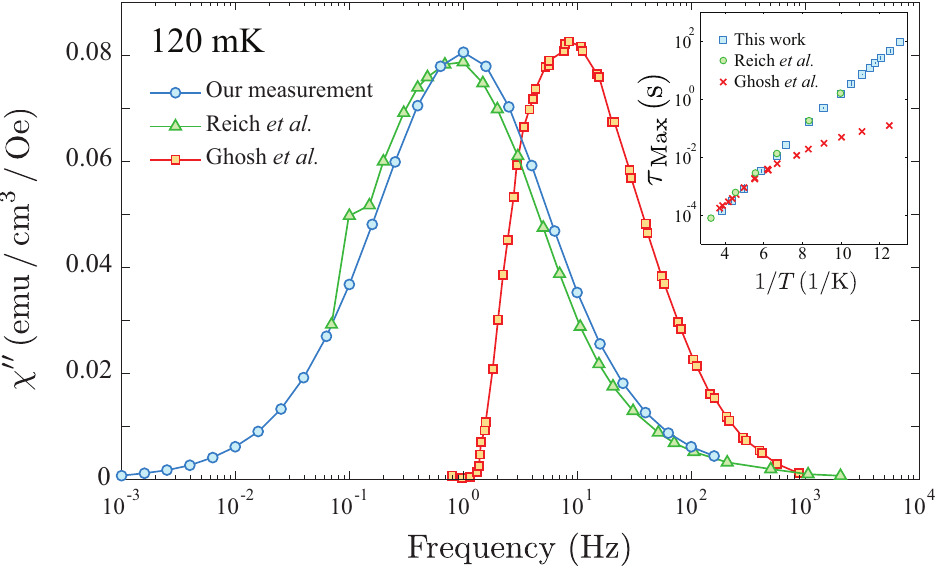}
\caption{(color online) Comparison of absorption spectra measured at 120 mK, showing good agreement between that measured in this work and by Reich \emph{et al.}\cite{Reich1990} but anomalous results from Ghosh \emph{et al.}\cite{Ghosh2002}  Note that \emph{no} arbitrary scaling of the vertical axis has been performed here.  The inset shows the maximum frequency of $\chi''(\omega)$ as a function of temperature for the same three measurements. 
\label{CompareReichGhosh}
}
\end{center}
\end{figure}

While the antiglass state was characterized by absorption spectra that narrow with decreasing $T$ and develop a strong asymmetry, we see clear and largely symmetric broadening down to 77 mK.  Furthermore, the temperature dependence of the peak frequencies $f_\mathrm{Max}(T)$ between our results and those of Ghosh \emph{et al.} are very different (see Fig.~\ref{CompareReichGhosh}).  The dc susceptibility reported to follow a $T^{-0.75}$ power law, is in fact found here to be much steeper with a temperature dependence closer to $T^{-1.2}$ for much of the temperature range of our data (see Fig.~\ref{TempScans}).  Finally, the specific heat presented in Ref.~\onlinecite{Ghosh2003} (part of which was presented previously in Ref.~\onlinecite{Reich1990}), exhibits two sharp peaks at 300 and 100 mK, and does not resemble our measured specific heat which has a single broad feature at around 120 mK (see Ref.~\onlinecite{Quilliam2007}).

We note that results by Reich \emph{et al.} and Ghosh \emph{et al.}, from the same research group and on the same sample,\cite{SilevitchComm} do not match in several ways.  A large part of the data of Reich \emph{et al.} agree quite well with our data.  We observe similar $\chi'(f) - i\chi''(f)$ spectra, with a comparably slow relaxation rate even well above $T_g$.  The peak frequencies, $f_\mathrm{Max}(T)$ are almost identical, and the dc limit of our data $\chi_{dc}(T)$ is quite close.  Reich \emph{et al.} show a slight broadening of the absorption spectra above 150 mK, which we do not observe, and our specific heat results disagree.  Otherwise, we obtain nearly the same results as Reich \emph{et al.}\cite{Reich1990} at a Ho concentration of $x=0.045$.  The behavior that clearly points to spin glass physics in our results is found, for the most part, at lower temperatures and frequencies than were obtained by Reich \emph{et al.}\cite{Reich1990}

\begin{figure}
\begin{center}
\includegraphics[width=3.25in,keepaspectratio=true]{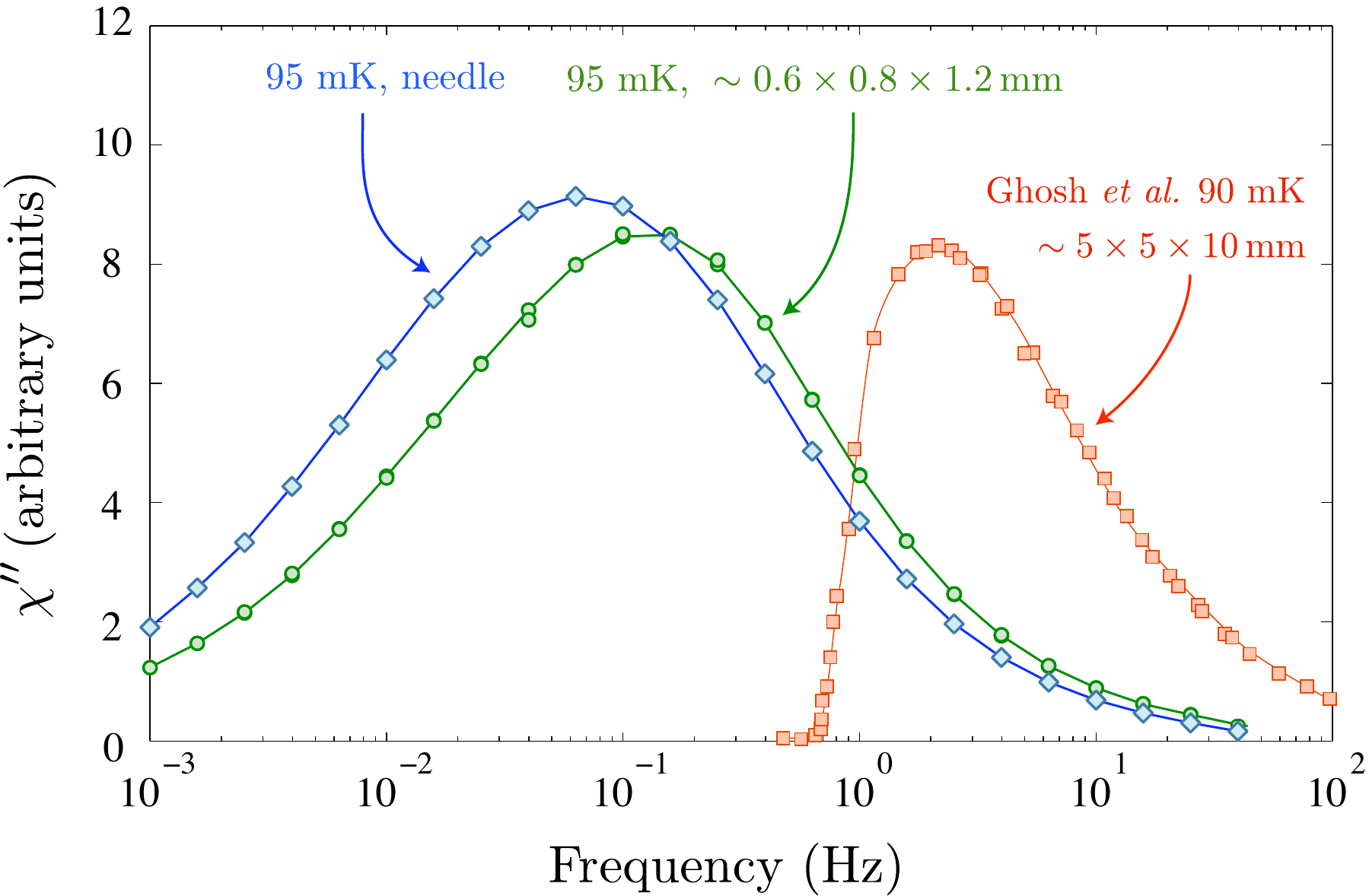}
\caption{(color online) Raw data, not corrected for demagnetization.  Our data, obtained at 95 mK, on two different sample geometries: a needle-like sample of dimensions $0.57\times 0.77\times 7.7$ mm$^3$ (blue diamonds) and a much less elongated sample of dimensions $0.57\times 0.77\times 1.2$ mm$^3$ (green circles).  Shown for comparison are data from Ghosh \emph{et al.}\cite{Ghosh2002}, taken at a similar temperature of 90 mK, on a sample with dimensions $5\times 5\times 10$ mm$^3$.  The difference between our two samples is well accounted for by the demagnetization effect, whereas the data of Ghosh \emph{et al.} are quite different, shifted to higher frequencies and quite assymetric.  Solid lines are guides to the eye and not theoretical fits.
\label{GeometryEffect}
}
\end{center}
\end{figure}

It is has been suggested that a possible explanation for the differences between the work of Ghosh \emph{et al.}\cite{Ghosh2002} and our results as previously published,\cite{Quilliam2008} could come from differences in sample geometry.  The measurements presented here were taken on two different sample geometries, one needle shaped and one less elongated.  The resulting spectra are in fact different, but this is likely a result of the demagnetization effect.  In the case of the contentious $x = 0.045$ sample, we have also measured a third geometry (very close to the geometry measured by Ghosh \emph{et al.}) to verify that our method of demagnetization correction is successful and that there are no strong changes in the true material $\chi(f)$ associated with the sample geometry.  As expected, a different shape of spectrum is observed for this third sample geometry, but the result is precisely accounted for by the demagnetization correction.  Furthermore, the differences between the results of Ghosh \emph{et al.} and those presented here are far too significant to be explained with the effects of sample geometry, as is demonstrated in Fig.~\ref{GeometryEffect}.

When comparing our results and those of Reich \emph{et al.} to those of Ghosh \emph{et al.}, we see that those of Ghosh \emph{et al.},\cite{Ghosh2002} are found to have higher characteristic frequencies and a shallower dc $\chi'$ as a function of temperature.  Since the characteristic frequency is a monotonically increasing function of $T$ and the dc susceptibility is monotonically decreasing in $T$, it appears that the measurements of Ghosh \emph{et al.} are essentially warmer than expected.  This may be explained through heating from larger oscillating fields, a problem of heat sinking, a lack of thermal equilibrium or perhaps a problem capturing quantitatively accurate susceptibility values at lower frequencies.  The sample studied by Ghosh and coworkers has a surface to volume ratio at least 10 times smaller than the samples studied here, possibly making it much more difficult to adequately heat sink to the dilution refrigerator.  The heating might also involve a complex mechanism, for example a phonon bottleneck, where heat dissipated into the magnetic moments is not easily released via phonons, effectively leading to poor heat sinking at low $T$ and strongly nonlinear heating effects.  The modified shape of the absorption spectra may be a result of variable amounts of heat dissipation in the sample at different frequencies of measurement, owing to the strong frequency dependence of $\chi''$.  This distortion of the spectra, which occurs predominantly on the low-frequency side, seems to lead to an inaccurate measure of $\tau_\mathrm{Max}(T)$ and an apparent narrowing of the spectra with lower $T$.  These ideas remain largely conjecture, however, as we have not been able to simulate or experimentally reproduce the results of Ghosh \emph{et al.} by introducing heating to the sample.

\subsection{Nonlinear susceptibility}

A third group of researchers, J\"{o}nsson \emph{et al.}, have also measured the bulk susceptibility of dilute LiHo$_x$Y$_{1-x}$F$_4$ single crystals.\cite{Jonsson2007}  These experiments involved linearly sweeping the external magnetic field and observing the resulting magnetization with a SQUID.  Then, over a certain range of magnetic field, they perform a fit of the form $M(H) = \chi_1H - \chi_3H^3$, in order to extract the linear and nonlinear susceptibility.  Their linear susceptibility result is shown in Figure~\ref{TempScans}, and is found to match well with ours down to a temperature of around 150 mK.\cite{JonssonDemag}  Although Jonsson \emph{et al.} are sweeping the magnetic field quite slowly, there is guaranteed to be some loss of equilibrium below a certain point, which appears to be occuring below 150 mK or so.  Because of the different measurement techniques employed, it is difficult to directly compare our results, particularly since the measurement of Jonsson \emph{et al.} does not operate at a unique or well defined frequency.  We can get a very approximate idea of what frequency of measurement is being employed through $f \sim (dH/dt)/ \Delta H$, where $\Delta H$ is the range of fields over which the data are analyzed and $(dH/dt)$ is the sweep rate employed.   Given the numbers quoted in Ref.~\onlinecite{Jonsson2007}, that frequency is, at the lowest, $\sim 40$ mHz.  While indeed quite a low frequency by most standards, it can be seen from Figure~\ref{TempScans}, that one should see a freezing temperature of around 110 mK.  

Thus, our results do not seem to be in disagreement with the data of Jonsson \emph{et al.}, but rather disprove the idea that their measurements are in the dc limit.  From the nonlinear susceptibility, $\chi_3$, they have found a rather broad peak, and have concluded that that feature is not sharp enough to represent a spin-glass transition.  They have argued, on the same basis, that there is not a spin-glass transition in an $x = 0.167$ sample either, which has opened up a debate regarding a part of the phase diagram previously thought to be well understood.\cite{AnconnaTorres2008,Jonsson2008}  It is true that the appearance of a peak in $\chi_3$, not accompanied by a scaling analysis, is insufficient evidence of a spin-glass transition.  

However, based on our dynamical measurements, we believe that the dynamics are rounding this peak and obscuring the glass transition.  While the critical exponent for $\chi_3$ is $\gamma\simeq 3$ in most spin glass systems, indeed leading to a sharp divergence, the dynamical critical exponent is much larger: $z\nu \simeq 8$.  Thus the dynamical component will always ``win'' at some point and eventually round the peak in $\chi_3$.  How far away from the true $T_g$ this occurs is dependent on $\tau_0$ and the frequency of measurement employed.  Thus for materials with a small $\tau_0$, something close to a divergence in $\chi_3$ may easily be obtained, but for systems with larger $\tau_0$, closer attention must be paid to the dynamics.  A measurement of $\chi_3$ at low $x$ in the LiHo$_x$Y$_{1-x}$F$_4$ series, could be a valuable future measurement and a definitive proof of spin glass physics, but caution is warranted, as the extreme values of $\tau_0$ should make it extremely difficult to obtain a sharp peak in $\chi_3$ and a careful dynamical scaling analysis would likely be necessary.

{\bf $\mu$SR experiments} -- Recently, Rodriguez \emph{et al.}\cite{Rodriguez2006,Rodriguez2010} have performed $\mu$SR experiments on the same samples of LiHo$_x$Y$_{1-x}$F$_4$ that were used for the work presented here.  Although they do not find any significant change in behavior between higher concentration samples ($x=0.12$ and $x=0.25$) and concentrations around $x = 0.045$ that would be suggestive of an antiglass state, neither do they obtain results that are typical of spin glass physics.  In fact, the $\mu$SR experiments show a temperature-independent fluctuation rate below at least 200 mK that is faster than the lower limit of the frequency window of $\mu$SR measurements.   This is reminiscent of the persistent spin dynamics (PSDs) that are observed in many geometrically frustrated magnets, including the pyrochlore spin glass Tb$_2$Mo$_2$O$_7$.\cite{Dunsiger1996}  

It is quite difficult to reconcile the idea of persistent spin dynamics exhibiting fluctuation rates ranging from 0.75 to 20 MHz with the sharply temperature-dependent dynamics reaching frequencies of less than $1\times 10^{-3}$ Hz, seen with susceptibility measurements in the same temperature range.  In several other magnetic systems, the microscopic fluctuation rate as determined with $\mu$SR has been found to be much faster than the characteristic frequency obtained from bulk ac susceptibility measurements.\cite{Dunsiger1996,Lago2007,Gardner2010}  Included is the dipolar spin ice material Dy$_2$Ti$_2$O$_7$, in which the Dy$^{3+}$ ions also carry Ising moments.  There, a qualitatively similar temperature dependence of the dynamics is observed in both $\mu$SR and in ac susceptibility, but there is an overall mismatch of time scales, many orders of magnitude in size.\cite{Lago2007}  In LiHo$_x$Y$_{1-x}$F$_4$, we do see some commonalities between the $\mu$SR and $\chi''(f)$ data:  there is a significant drop in the low-temperature fluctuation rate, $\nu_0$, as $x$ is lowered, coincident with our observation of an increase in $\tau_0$.  Rodriguez \emph{et al.}\cite{Rodriguez2010} also draw attention to the importance of the nuclear hyperfine interaction in this system with the observation that the plateau of the fluctuation rate has an onset near 300 mK, not so distant from the 200 mK energy scale associated with $\mathcal{H}_{HF}$.

\begin{figure*}
\begin{center}
\includegraphics[width=5.55in,keepaspectratio=true]{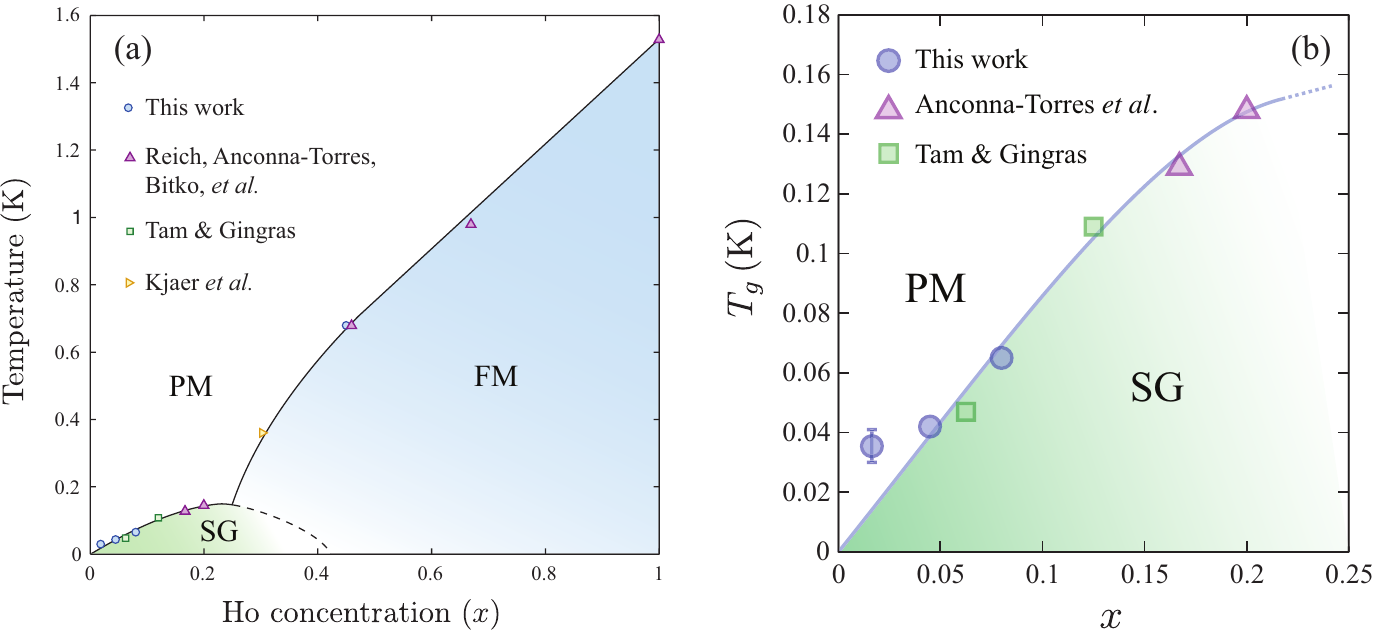}
\caption{(color online) Current understanding of the phase diagram of LiHo$_x$Y$_{1-x}$F$_4$ as a function of concentration $x$ and temperature $T$.  Ferromagnetic transition temperatures are obtained at $x = 1$ by Bitko \emph{et al.}\cite{Bitko1996}, at $x=0.44$ and $x=0.67$ by Reich \emph{et al.}\cite{Reich1990}, at $x=0.45$ by Quilliam \emph{et al.} and at $x=0.30$ by Kjaer \emph{et al.}\cite{Kjaer1989}.  Spin glass transition temperatures are determined experimentally in this work (blue symbols) and by Anconna-Torres \emph{et al.}\cite{AnconnaTorres2008} (purple triangles) and with Monte Carlo by Tam and Gingras\cite{Tam2009} (green squares).  A rather large error bar is shown with the poorly defined data point at $x=0.018$.  The solid lines are guides to the eye rather than theoretical models.  In part (a), a region of reentrance has been drawn, as is often observed in other such systems.\cite{Reentrance}  However, this occurrence has not yet been observed in this system.
\label{PhaseDiagrams}
}
\end{center}
\end{figure*}

\subsection{Theory}

With some exceptions, our results are very close to theoretical work on the ideal dilute dipolar Ising model.  The glass temperatures obtained here for $x=0.045$ and $x=0.08$ are well matched to those determined in Monte Carlo work of Tam and Gingras,\cite{Tam2009} as can be seen in Fig.~\ref{PhaseDiagrams}(b).  Although the theoretical work determines $T_g$ for the concentrations $x = 0.0625$ and $x=0.125$, we can easily interpolate between those points and find adequate agreement between experiment and theory.  The 1.8\% sample shows an anomalously high $T_g$ when compared to theory, however.  In section V, we have suggested two possible nuclear hyperfine effects that may explain this discrepancy.  Biltmo and Henelius,~\cite{Biltmo2008} have also performed Monte Carlo simulations on the ideal model, notably with $x=0.045$, and have computed several physical properties, including the dc susceptibility and specific heat.  The dc limit of our ac susceptibility measurements are in fact very much in line with those simulations, as shown in Fig.~\ref{TempScans}.  The specific heat, however, is found in theory to exhibit a broad peak that shifts in temperature with $x$.  The measured specific heat exhibits a similarly broad peak, but one that does not change with $x$.  Thus this discrepancy between experiment and theory remains an important question to issue to resolve.

\section{Conclusions}

To summarize, we have shown that a dynamical scaling analysis of ac susceptibility measurements taken on low-$x$ samples of LiHo$_x$Y$_{1-x}$F$_4$ gives strong evidence of finite-temperature spin-glass transitions.  Furthermore, these glass transition temperatures, $T_g$, are well matched to theory\cite{Tam2009} as can be seen in the phase diagrams plotted in Fig.~\ref{PhaseDiagrams}.  There remains some ambiguity regarding the glass temperature $T_g$ of the 1.8\% sample and the specific heat of all the samples that remains to be fully understood.  We have tentatively proposed that these anomalies in our understanding of this system may be the result of the complicated role of the nuclear hyperfine interaction on thermodynamic and dynamic quantities.  Observation of the antiglass phase,\cite{Reich1987,Reich1990,Ghosh2002,Ghosh2003} conflicts with the behavior that we have deduced with our ac susceptibility and specific heat experiments.  While tentative explanations for these discrepancies have been suggested, a final and quantitative explanation remains to be discovered.  Given recent theoretical conclusions,\cite{Tam2009} inconsistencies in the antiglass picture,\cite{Reich1990,Ghosh2002} and the clear appearance of very slow dynamics in ac susceptibility with conventional scaling exponents, $z\nu$, we feel, nonetheless, that there should remain little doubt that LiHo$_x$Y$_{1-x}$F$_4$ does indeed freeze as a spin glass at low $x$.  

Most interestingly, our results show compelling evidence for the very important effect of nuclear hyperfine interactions on the dynamics of the system.  Typically nuclear moments are coupled weakly and have very little impact on electronic magnetism, although they can routinely serve as a means to study the electronic physics of a system through NMR or M\"{o}ssbauer experiments, for example.  This system represents a rare case where the nuclear moments are hugely influential, in blocking the effects of internal, fluctuating transverse fields hence appreciably slowing the dynamics of the system.  The importance of the nuclear hyperfine interaction has also been observed or theoretically proposed under the influence of an external transverse magnetic field, when studying the effects of quantum fluctuations on the system.\cite{Bitko1996,Wu1993,Schechter2005,Schechter2008b}  These unusually slow dynamics have likely caused some of the confusion regarding the existence of a spin glass state at $x=0.167$,\cite{Jonsson2007,AnconnaTorres2008,Jonsson2008} where experiments with an \emph{arbitrary} time scale were used to argue for or against a finite $T_g$.  The results obtained here show the importance of exploring scaling relations and a careful extrapolation to the dc limit, particularly when studying glassy systems.  Precise calculations of the tunneling rates of the Ho$^{3+}$ moments are beyond the scope of this work, but it is hoped that such an exercise may be pursued in the future.  Such calculations may also have important implications for the dynamics of other rare-earth Ising systems, including those of the recently discovered magnetic monopole excitations in dipolar spin ice materials.\cite{Jaubert2009,Quilliam2011HTO}


\begin{acknowledgments}
We acknowledge many very informative discussions with M.~J.~P.~Gingras, P.~Henelius, M.~Schechter, L.~R.~Yaraskavitch,   S.~M.~A.~Tabei, K.-M.~Tam, G.~M.~Luke, J.~Rodriguez, and R.~W.~Hill and thank C.~G.~A.~Mugford for his many contributions to the experimental apparatus employed in this work. We acknowledge M.~J.~P.~Gingras in particular for a detailed reading of the manuscript.  Funding for this research was provided through grants from NSERC, CRC, CFI, MMO and The Research Corporation.
\end{acknowledgments}

\end{document}